\documentclass[useAMS,usenatbib,usegraphicx]{mn2e}
\usepackage{lscape}
\usepackage{rotating}

\def\c{{\sl Chandra}}
\def\xmm{{\sl XMM-Newton}}
\def\rosat{{\sl ROSAT}}
\def\asca{{\sl ASCA}}

\def\sax{{\sl BeppoSAX}}
\def\sirtf{{\sl SIRTF}}
\def\subaru{{\sl SUBARU}}

\def\ltsim{\mathrel{\hbox{\rlap{\hbox{\lower4pt\hbox{$\sim$}}}\hbox{$<$}}}}
\def\gtsim{\mathrel{\hbox{\rlap{\hbox{\lower4pt\hbox{$\sim$}}}\hbox{$>$}}}}
\def\ergpspsqcm{ erg s$^{-1}$ cm$^{-2}$}
\def\ergps{ erg s$^{-1}$}

\def\micron{$\mu$m}
\def\p{$\pm$}

\def\zp{$z_{\rm phot}$}
\def\zphot{$z_{\rm phot}$}
\def\zs{$z_{\rm spec}$}
\def\zspec{$z_{\rm spec}$}

\def\av{$A_{\rm V}$}
\def\k{{\sl K}}
\def\h{{\sl H}}
\def\j{{\sl J}}
\def\lx{$L_{\rm X}$}

\def\lognh{log$N_{\rm H}$}
\def\nh{$N_{\rm H}$}
\def\l{$\lambda$}

\begin{document}

\title{ Powerful obscured AGN among X-ray hard, optically-dim serendipitous {\em Chandra} sources }

\author[P. Gandhi et al.]
{\parbox[]{6.in} {P. Gandhi$^{1,2}$, C. S. Crawford$^1$, A. C. Fabian$^1$, R. M. Johnstone$^1$\\
\footnotesize
1. Institute of Astronomy, Madingley Road, Cambridge CB3 0HA\\
2. European Southern Observatory, Alonso de Cordova 3107, Santiago 19, Chile\\
}}

\maketitle
\begin{abstract}
We present a small sample of \c\ X-ray sources selected from the fields of ACIS observations which probe fluxes around the break in the hard band source counts.
The targets of these fields include 9 nearby galaxy clusters, 1 distant cluster and 2 powerful radio galaxy fields. The follow-up of this serendipitous sample was biased towards X-ray hard and optically-dim sources mostly not seen on the Digitized Sky Survey; for these, we present X-ray fluxes, optical and near-infrared photometry leading to 51 photometric redshifts in all and 18 independently measured spectroscopic redshifts. Few sources are associated with the target fields themselves. Fifty-six of 58 sources imaged in the \k-band are detected at \k$\ltsim$20 with \k$_{\rm median}$=18, and of these, 38 have hard X-ray count ratios and 24 of these are significantly hard with most of the counts emerging about 2 keV. 
We find that almost all are AGN hosted in massive early-type host galaxies with a photometric redshift distribution peaking at $z$$\sim$1.
Two type 2 quasars with intrinsic X-ray luminosity $L$$\gtsim$10$^{45}$ \ergps, Fe K$\alpha$ emission lines and absorbing column density \nh$>$10$^{23}$ cm$^{-2}$ -- and \nh$\gtsim$10$^{24}$ cm$^{-2}$ in one case -- are discussed in detail; the sample contains at least 12 potential type 2 quasars in all. We discuss various detection strategies for type 2 quasars and calculate their inferred space density. 
This combines and extends a number of results from subsamples already published by us.
\end{abstract}


\section{Introduction}

The great effort of following-up the X-ray sources which power the X-ray background (XRB) is now beginning to yield results. Deep, pencil-beam surveys in particular are compiling complete samples to ever-fainter fluxes, and provide important constraints on accretion and even star-formation to high redshifts \citep{alexander_2Ms, giacconi02, hasinger01, brandt02}.

While there may be several per cent of the 2--10 keV XRB which still remains unresolved \citep{moretti03, miyajigriffiths02}, the bulk of the background itself was resolved to within the uncertainty of its known absolute level in the first deep targeted 100 ks exposure with \c\ \citep{mushotzky00}. The major contribution to the XRB flux emerges at fluxes within \p1 dex of the break in the source counts at $\sim 10^{-14}$ \ergpspsqcm\ (e.g., Fig~3 of \citealt{cowie02}). Though this is two orders of magnitude brighter than the faintest fluxes reached in Ms \c\ exposures, previous hard X-ray missions such as \sax\ and \asca\ did not possess the adequate combination of sensitivity and resolution to probe it; this flux regime thus remains largely unexplored. A number of wide-area X-ray surveys are currently underway with \c\ and \xmm\ \citep{champ, watson01, hellas2xmm, sexsi}; these have recently started to deliver results and will probe complete samples over large areas at this flux over the next few years. These studies are essential to bridge the gap between ultra-deep pencil-beam surveys and all-sky shallow surveys.

In the mean time, we have been carrying out a selective study of point X-ray sources in the fields of a number of nearby ($z\sim 0.1-0.5$) galaxy clusters and powerful radio galaxies that have been observed with the ACIS instrument aboard \c. In contrast to the complete samples being compiled by other groups, we have aimed to study the sources with hard X-ray count ratios (presumably obscured) and optically-dim counterparts -- exactly the kind which would have been missed in previous X-ray and optical AGN surveys. The main motivation for adopting this strategy has been to concentrate on these interesting sources within the limited amount of ground-based telescope time available, and we have been able to discover and publish results on several type 2 quasars [For the sources presented herein, we use the working definition of a type 2 quasar as a source with an intrinsic, rest-frame X-ray luminosity in the 0.5--7 keV band $L_{0.5-7}\ge 10^{44}$ erg s$^{-1}$ and showing indications of significant X-ray absorption above that expected due to the Galaxy alone]. 
The \c\ exposure times are typically in the range $10-30$ ks and straddle the break in the hard band source counts, thus maximizing source detection per unit exposure.
The main results from a number of our publications can be summarized as follows (\citealt{c01,c02} [hereafter C01, C02 respectively]; \citealt{g02} [hereafter G02]):\\
$\bullet$ In the field of Abell~2390, we found 31 X-ray point sources to a 0.5--7 keV flux limit of $\sim 7\times 10^{-15}$ \ergpspsqcm\ over the entire ACIS-S field of view.\\
$\bullet$ One-third of these sources have hard X-ray count ratios, i.e. the soft band (0.5--2 keV) to hard band (2--7 keV) count ratio $<$ 2.0. For reference, on ACIS-S3, this ratio is measured as 5.7 and 3.4 for a power-law spectrum ($F_E\propto E^{-\alpha}$) with energy-index $\alpha=1$, at $z=1$, absorbed by columns of $0$ and $10^{22}$ cm$^{-2}$ respectively local to the AGN and a moderate galactic column of $2\times 10^{20}$ cm$^{-2}$.\\
$\bullet$ These X-ray point sources have predominantly AGN-like optical spectra: of 15 X-ray sources in the field of A\,2390, we found 5 broad-line and 7 narrow-line AGN. One source showed a galaxy-like spectrum with no distinct line emission, while two sources were characterized as stars.\\
$\bullet$ The optical magnitudes of the AGN and galaxy-like sources cover a wide range from $R\sim 20-24$, while the same sources can usually be detected with near-IR imaging to $K\sim 20$ on a 4m telescope.\\
$\bullet$ Photometric redshifts derived from optical/near-infrared photometry work well for X-ray selected AGN with hard X-ray count ratios. These are sources in which host galaxy light is the principal component, while AGN light is scattered out of the line-of-sight due to large optical-depths of obscuring gas and associated dust.\\
$\bullet$ Near-infrared spectroscopy of several hard X-ray sources revealed flat continua with little or no line emission; the limiting line fluxes and equivalent widths were deep enough to infer the presence of dust responsible for scattering line photons in more than half the sources observed.\\
$\bullet$ We have identified at least two bona-fide type 2 quasars with local obscuring gas column density \nh$\sim$2$\times$10$^{23}$ cm$^{-2}$: source A18 in the field of A\,2390 with intrinsic X-ray luminosity $L_{\rm 2-10\ keV}$$>$10$^{45}$ erg s$^{-1}$; and source A15 in the same field, inferred to have a large intrinsic ultraviolet luminosity $L_{\rm UV}$$>$10$^{45}$ erg s$^{-1}$ being absorbed by dust, based on radiative transfer modeling of detections by the Infrared Space Observatory.\\

\noindent
In this paper, we extend our sample to include serendipitous X-ray sources in a larger number of \c\ fields, still aiming to identify intrinsically powerful and highly obscured AGN. In any medium-deep X-ray observation, a large fraction of detected sources are unobscured type 1 AGN. These are relatively easy to follow-up due to their brightness in the soft band and are likely to possess broad optical emission lines, as previous studies have shown \citep{mi00, lehmann01}. We do not concentrate on these sources, but include some for comparison. 

To briefly summarize our follow-up strategy: we preferentially select X-ray sources with the hardest X-ray count ratios in each field and cross-correlate with optical catalogues of cluster fields drawn from various telescope archives. In case such archives are unavailable, we search for a counterpart in the Digitized Sky Survey (DSS) catalogue. We define optically-dim sources as ones with faint or no counterparts on the DSS (limiting magnitude $B\sim 22.5$, $R\sim 21$). The X-ray hard, optically-dim sources are then imaged in the near-infrared and photometric redshifts are determined in as many bands as possible. From a total X-ray sample of more than 300 sources, we are able to follow-up and derive redshifts for 57. We find a number of sources whose count ratios and inferred luminosities classify them as type 2 quasars selected in X-rays.

Several detailed studies of individual type 2 quasars have emerged from current \c\ and \xmm\ surveys \citep{stern02, willott03, norman02} and in the past with \rosat, \sax\ and \asca\ \citep{almaini95,franceschini00_iras09,nakanishi00}. It is likely that the population of Seyferts and quasars being discovered with the current generation of X-ray telescopes are radio-quiet (\citealt{barger01}) [Radio-loud obscured AGN are thought to be $\sim 10$ times less numerous and have been associated for a long time with the population of radio galaxies (e.g., \citealt{urrypadovani95})]. The study of such sources is important not only in testing the robustness and evolution of AGN unification schemes, but also in the fact that they are undergoing extremely powerful accretion activity and are probably at the epoch of the most rapid growth of their central black holes \citep[e.g., ][]{f98}. New observations \citep{cowie03, hasinger03} suggest that the evolution of AGN may be luminosity-dependent, in which case determining the evolution of the ({\em intrinsically}) bright-end of the X-ray luminosity function is clearly an important goal of X-ray surveys. With our current sample, we are in a position to estimate the number density of X-ray selected type 2 quasars, and to compare with other findings.

Twenty-five sources presented herein have no previously-reported detections, while three others have published redshifts of single-filter photometry in the literature by other authors (\S~\ref{sec:zphot}). This paper combines and extends results from our earlier works (C01, C02, G02), roughly doubling our already-published sample. We also discuss an \xmm\ X-ray spectrum for the source A2390\_18, previously studied by us with \c. We assume $H_0$=70 km s$^{-1}$ Mpc$^{-1}$, $\Omega_{\rm M}=0.3$ and $\Omega_{\Lambda}=0.7$ throughout (referred to as the standard cosmology), and all quoted magnitudes are in the Vega system.

\section{Sample Selection and Observation}

\subsection{X-ray observations}
\label{sec:xray}

As part of the guaranteed time of one of us (ACF) as well as guest observer time on \c, a number of nearby ($z\sim 0.02-0.4$) galaxy clusters have been observed with the ACIS instrument. 
The exposure times are typically in the range 10--30 ks, which corresponds to an ACIS-S3 (back-illuminated) flux limit of $7 - 4\times 10^{-15}$ and an ACIS-I (front-illuminated) limit of $10 - 6\times 10^{-15}$ \ergpspsqcm\ over the 0.5--7 keV range, assuming a $\Gamma=1.4$ power-law photon-index\footnote{The photon-index is the slope of the power-law photon flux density: $N_E \propto E^{-\Gamma}$}. 
Additionally, we have studied the fields of the powerful radio galaxy B2~0902 ($z=3.395$), the distant cluster 3C~294 ($z=1.786$), and the observation of the cluster MS~2137-2353 has been drawn from the \c\ X-ray Centre\footnote{http://cxc.harvard.edu/cda/} archive. In a longer follow-up 70~ks X-ray exposure of 3C~294, we reach a limiting S3 flux of $\sim 2\times 10^{-15}$ \ergpspsqcm\ for background source detection (\citealt{3c294longest} have published an even longer 200~ks exposure of this field). 
Table~\ref{tab:fieldlist} lists all the fields in the present analysis.

The data analysis and source detection procedure has been described for a number of clusters in C01, C02 and G02, and is identical for the rest of the fields. Here we review it briefly. The \c\ Interactive Analysis of Observations (CIAO) software was used throughout, with the latest software version and calibration files available at the time of observation. WAVDETECT was used for source detection in each of three bands: 0.5--2 keV (Soft; S); 2--7 keV (Hard; H) and 0.5--7 keV (Total). A full range of wavelet scales using the $\sqrt{2}$ sequence was used to detect different-sized sources. The exposure times stated in Table~\ref{tab:fieldlist} are effective exposure times, after removal of background flaring. All fields were exposure-map corrected prior to running WAVDETECT. Source extraction was performed on individual exposures -- fields observed more than once were not combined (for spectral analysis, the flux from all exposures was analysed jointly with the appropriate response matrices). For a discussion of source flux variability between two exposures in the field of Abell~2390, see C02. 

Sources with fewer than 10 net counts and those within 20 arcsec of the chip outer edges were discarded. Each source was examined by eye, in order to discard any that could be due to the prominent stripes apparent in the background of some of the chips. The number of sources detected in the full 0.5--7 keV band in each field are listed in Table~\ref{tab:xraysample} (second column). The procedure for selecting sources for follow-up is described further on in this section (\S~\ref{sec:selection}), while the catalogue with X-ray coordinates, fluxes and hardness ratios is presented in Table~\ref{tab:xraycat} on p.~\pageref{tab:xraycat}ff. Thumbnail X-ray images of all sources are available in Appendix A of the electronic version of this paper\footnote{and http://www-xray.ast.cam.ac.uk/$^\sim$pg/xrbsources/imsp/}.


It is difficult to accurately estimate limiting fluxes and source densities within an $\sim 1$ arcmin radius region of the cluster cores, due to the high background from the cluster itself and lensing effects such as magnification and gravitational depletion. Typically, the cluster fields observations contain one X-ray point source within such as region (including A15 and A18 lensed by Abell~2390; C02). Under the assumption that these are all background sources at $z=1$, strong lensing amplification factors are inferred to lie between 1.2--3, depending on the mass model of the cluster and the exact distance of the source from the core.

While our primary aim is not to compile a complete X-ray sample, we note that the average source density found by simply dividing the total number of detections by the sky area covered by \c, assuming a uniform limiting flux per observation is fully consistent with other studies \citep[e.g., ][]{mushotzky00}. We find approximately 313\p85, 334\p90 and 835\p211 sources deg$^{-2}$ to 2--10 keV flux limits of $\sim 1\times 10^{-14}, 7\times 10^{-15}$ and $2\times 10^{-15}$ \ergpspsqcm\ respectively. The 1-$\sigma$ poisson errors thus also reflect the combined variations in the detectability of sources from chip-to-chip and with off-axis angle, as well as \lq cosmic variance\rq\ from field-to-field.\\

\subsection{Optical counterparts}
\label{sec:optical}

Deeper (than the DSS) identifications of or detection limits to optical counterparts can often be obtained from the rich resources of archival data maintained by many telescopes. This is especially true for sources which lie within a few arcmin of galaxy cluster cores, likely to have been observed in the past. We have been able to make use of such data from a number of telescopes: the Issac Newton Telescope (INT) and the William Herschel Telescope (WHT) on La Palma, the Canada-France-Hawaii Telescope (CFHT), the Anglo-Australian Telescope (AAT) and in a few cases, the Hubble Space Telescope (HST). The field of view of the respective instruments used determines the number of \c\ sources that can be imaged (the Wide Field Camera [WFC] on the INT, with each chip having a large field of $22\times 11$ arcmin$^2$, is especially useful in this regard), and sources detected far from the cluster core (e.g., on ACIS chips I2 or I3) often have only DSS identifications or limits.

All the (science and calibration) data were requested and down-loaded from the respective telescope archive web-sites. We obtained data in as many filters as possible -- this was a mixture of filters in the $B$, $R$ and $I$ regime, and also included $U$ and $V$ in a few cases. Calibration involved bias-subtraction and flat-fielding using archival frames created on the night of observation. Fringing was removed from $I$-band data by generating a master fringe frame from typically 6--8 offset frames, where possible.

In addition, several of the follow-up datasets were obtained through successful service time applications. Finally, some calibrated proprietary optical data for the fields of A\,2199 and A\,1795 was also obtained from Neil Trentham. Table~\ref{tab:optical} gives details of the deeper optical datasets used.

Seeing full-width at half-maximum (FWHM) diameters for the
optical observations were typically 1~arcsec or more. In many cases, this led to unreliable morphological classification from the optical data alone, especially for faint sources or those with close neighbours.


\subsection{Photometry}
\label{sec:photometry}

Optical (and near-infrared; NIR) magnitudes were computed using the SExtractor package \citep{sextractor}. These
are Kron magnitudes (with a Kron scale of 2.5; \citealt{kron80}) for isolated
sources and seeing-corrected isophotal magnitudes
for blended objects. We also used the PHOT task in IRAF as a cross-check on the photometry for a few sources. The magnitudes that we measure are an estimate of the {\sl total} flux of each source\footnote{With the Kron scale used, we can expect to encompass at least 94 per cent of the source flux (\citealt{kron80}; also SExtractor V1.0a users guide, E. Bertin)}. In some cases (e.g., for sources at the edge of the field-of-view or for sources with large seeing variations in different bands), we estimate the magnitude in a smaller aperture, but keep a consistent aperture for all bands in order to provide a uniform estimate of the flux for photometric redshift fitting (discussed later). In a few cases (e.g., $I$-band data without proper fringe maps), apertures were selected by hand and background maps were inspected for consistency.

Zeropoints were determined in all cases (unless otherwise stated in the catalogue tables) from photometric standard star observations taken close in time to the science targets. For the NIR data described in \S~\ref{sec:nir}, typically 3 stars (usually from the faint standards catalogue in the MKO filter set at UKIRT and the ISAAC filter set at the VLT; \citealt{ukirt_faintstandards}) were observed
over the course of the night and the zeropoint in each filter was
found to be constant to within 0.02 magnitudes on photometric nights. It has been reported that the {\sl
jitter} pipeline used at ISAAC may underestimate the brightness of sources in the
\k-band due to biasing of the local background in jittered images. We
corrected for this by a simple prescription of increasing all the
fluxes obtained at the VLT by 10 percent. Refer to \citet{iovino01} for more
details.

We had less (or no) control over the choice of optical data observing conditions taken in service mode or drawn from the archive. DSS magnitude measurements used a smooth extension of the first generation flux calibration by the Catalogs and Surveys Branch\footnote{http://www-gsss.stsci.edu/; based on observations by Lasker et al. (1988)} to fainter fluxes. \citet{landolt92} standard star observations were used for calibrations of all the deeper fields. Each Landolt field contains a number of stars (at least 5) which can be used for photometry. Calibration of the extinction, however, requires observations of standard stars at a variety of airmasses, and such observations were not always available in the archive. In these cases, we consulted archival observing night logs and used standard extinction measurements. This could be an additional source of error and we implicitly account for it during photometric-redshift fitting by assuming a minimum systematic error of between 0.05--0.1 magnitudes.

Upper-limits for
non-detected objects were defined as the flux corresponding to 3 times
the background sky RMS in a 3-arcsec diameter aperture close to the source
location. Such an aperture size is typical of the Kron apertures for
the fainter of the detected sources. 


\subsection{Source selection}
\label{sec:selection}

Sources with an X-ray count ratio S/H$<$2.5 are consistent with power-law spectra having a photon-index harder than $\Gamma=1.4$ (the slope typically associated with X-ray background spectrum at less than 10 keV; e.g., \citealt{marshall80}) at all redshifts. We define this S/H ratio as the limit for \lq hard\rq\ sources. However, the errors on this ratio can be large and we thus choose a limit of S/H$\le$1.5 for \lq very hard\rq\ sources; these are typically hard (i.e. S/H$\le$2.5) even at 95 per cent confidence. Table~\ref{tab:xraysample} lists the size of the total detected \c\ sample of X-ray sources in each field. In the third column of the same table, we list the number and fraction of sources which have a S/H ratio less than with the average value of 2. 
The numbers show that these sources comprise between 20 and 50 per cent of the total X-ray sample at the flux levels that we probe. 



Several \lq soft\rq\ sources, defined as those with S/H$>$2.5, are included for comparison. These are typically sources which just happened to lie in the large UFTI field-of-view while imaging other primary X-ray sources. The fact that they are labelled \lq soft\rq\ does not preclude the possibility of their being obscured AGN at high redshift, due to the positive $k$-correction of hard counts into the soft band.

Of the hard sources, we preferentially target for NIR imaging the ones that are dim or invisible on the DSS (hereafter referred to as optically-dim). By analyzing the deeper optical images, we find that optically-dim sources in 10--20 ks ACIS exposures cover a wide range in magnitude: from $B\sim 22$ and $I\sim 19$ at the bright end to $B\gtsim 24$ and $I\gtsim 23$ at the faint end. Most of these are detected in the NIR at 16.5$\ltsim$$K$$\ltsim$19. Faint sources in deeper (35~ks) \c\ exposures can have optical limits of $B$$>$25.5, $I$$>$24 and $K$$>$20.

Our \lq best\rq\ targets are thus the hard sources that are also optically-dim
and we estimate these to be about 50 per cent of the
hard X-ray serendipitous sample. Selection was also governed by real-time observing conditions however, and we were forced to include a few sources (some of which are very
hard) which have brighter optical counterparts on the DSS. The last column in Table~\ref{tab:xraysample} lists the final number of sources for which we were able to obtain NIR photometry. The resulting magnitudes of all selected sources are presented in the optical-NIR catalogue in Table~\ref{tab:mags} on p.\pageref{tab:mags}ff.

Selecting optically-dim sources automatically biases our sample against local or very low redshift AGN. As an example, an $M^*$ galaxy with $M_{\rm K}=-23.9$ \citep{gardner97} formed in a single burst at $z=10$ would become invisible in the DSS $B$-band ($B_{\rm lim}\sim 22.5$) only beyond $z\approx 0.3$ and in the $R$-band ($R_{\rm lim}\sim 21$) beyond $z\approx 0.4$. 

Lastly, we note that having one of the primary selection criteria based on hardness ratio and then mixing samples from both ACIS-I and ACIS-S will lead to a bias due to the differing effective areas of the front- and back-illuminated CCDs. For instance, a source at $z=1$ with (log$N_{\rm H}$)$_{22}$ [in units of $10^{22}$ cm$^{-2}$]=1.6 will be classified as \lq hard\rq\ if observed on the S2 chip, while a source at the same redshift observed with the back-illuminated S3 must have (log$N_{\rm H}$)$_{22}>2.2$ to fit in the same category (see Fig. 2 of G02 for details and an illustration of the effect of varying hardness ratios with redshift and levels of obscuration). In any case, the poisson statistical errors on the observed counts are the dominant effect in our sample and we thus ignore differences between the two CCDs.

\subsection{Near-infrared imaging}
\label{sec:nir}

The selected sources were followed up in the near-infrared (NIR) with three separate instruments at the UKIRT and the VLT. 

\subsubsection{UFTI on UKIRT}

The UKIRT Fast-Track Imager (UFTI) at the United Kingdom InfraRed Telescope was the camera used to study the majority of objects in our sample. UFTI was used in ND~STARE mode, with standard read-out, and images were obtained in the jitter-mode (typically a 9-point grid with 10~arcsec offsets) over the full array with a 92~arcsec field-of-view at 0.091~arcsec/pixel. Dark current subtraction, flat-fielding, image registration and combination were all performed with the standard ORAC-DR software. Targets were centred on the UFTI frame unless two or more X-ray sources could be imaged in a single observation, in which case the aim-point was offset appropriately (the areal density of \c\ sources at our flux limits is $\sim$0.5--1 arcmin$^{-2}$; thus most UFTI fields contained one or two X-ray sources). Typical observing times were 60~s at each of the grid points, leading to a total exposure time of 540~s in each of the broad-band \j, \h\ and \k\ filters to limiting magnitudes (3$\sigma$ in a 3~arcsec diameter aperture) of $\sim$ 21.4, 20.7 and 20.0 respectively. A few interesting sources were specifically imaged for double or triple the integration times, while a few others (e.g., softer sources) happened to be imaged multiply during observations of neighbouring targets, and combination of these images leads to limiting magnitudes which are correspondingly fainter. 

The data were accumulated over approximately 8 nights of observing time during two separate observing runs in Aug 2000 and Jan 2001 (Table~\ref{tab:nirlog}). Typical seeing conditions ranged from 0.5--1.5 arcsec and all severely non-photometric data have been discarded from the final sample. Flux-calibration used observations of several standard stars in the same bands carried out during the course of each night, as described in \S~\ref{sec:photometry}.


\subsubsection{IRCAM3/TUFTI on UKIRT}

IRCAM3/TUFTI is an imaging camera with a scale of 0.081 arcsec/pixel,
and a total field of view of 20.8$\times$20.8 arcsec. Objects in two fields (Abell~1795 and IRAS09; Table~\ref{tab:nirlog}) were imaged in \j, \h\ and \k\ to approximate limiting magnitudes 21.6, 20.8 and 20.1 respectively on the nights of 24--25 Feb 2001. The purpose of these data was to supplement NIR spectroscopic observations of hard sources with CGS4 on UKIRT (C01; see that paper for more details). Imaging observations were carried out in jitter-mode (each grid point separated by 6 arcsec) and data reduction was carried out by the TUFTI tasks in ORACDR.

\subsubsection{ISAAC on the VLT}

Near infrared images of selected sources in the fields of three galaxy clusters observable from the southern hemisphere -- Abell~1835, Abell~2204 and MS\,2137.3-2353 -- 
were obtained using the imaging spectrograph ISAAC on VLT with a 2.5$\times$2.5~arcmin field at 0.148 arcsec/pixel. These were primarily pre-imaging fields for deep near-infrared spectroscopy carried out with ISAAC (G02).
We observed through the \j, \h\ and \k s (hereafter
\k) filters on 2001 June 28 in seeing of $\sim$0.3 arcsec (Table~\ref{tab:nirlog}).
The total integration time was 600 s in \h\ and
\k, and 720--900 s in \j, with individual exposures of 10 s obtained
in jitter mode around a grid with offsets of $\sim$30 arcsec. Bad
pixel map creation, dark current subtraction and flat-field division
were carried out using the {\sl jitter} routine of the {\sl eclipse}
software package V4.0.4 \citep{devillard97}. In addition, {\sl jitter}
was used for background subtraction (using parameters suggested by
Iovino 2001) and combination of jittered frames.\\

\noindent
Of 58 sources imaged in the \k-band, we are able to detect 56. The two non-detected sources 
are in the \c\ exposure of Abell~963, and both have very large X-ray to \k-band flux ratios $F_{\rm X}:F_{\rm K}>10$ (discussed later). The deepest \k-band detections are at \k=20.7, while \k$_{\rm median}$=17.7. The majority of the sources have morphologies consistent with being extended (i.e., FWHM significantly more than the seeing on the night of observation). Table~\ref{tab:nirlog} provides a log of the near-infrared observations for all the imaging fields. \k-band thumbnail images of all sources are presented alongside the X-ray images in Appendix A of the electronic journal.


\subsection{Source Matching}

Our source matching procedure has been described in C01 and G02. Briefly, all NIR fields were cross-calibrated with the APM sky survey catalogue\footnote{http://www.ast.cam.ac.uk/$^{\sim}$mike/apmcat/} and/or the DSS to generate an astrometric solution for each. RMS errors for the solutions were less than or close to a pixel (of the order of the pixel scale) in all cases over large image regions.
In a few cases where astrometric solutions could not be attached to the NIR data (due to a paucity of enough bright sources, for example), a solution was attached to any deep optical archival dataset and this was used to determine offsets.

Any astrometric offset between a \c\ source and its corresponding NIR
counterpart is typically least ($< 2$ arcsec) for sources on the
ACIS-S3 chip 7, closest to the telescope aim-point. With off-axis PSF
degradation, this may increase to several arcsec on the other chips.
 Systematic errors are also introduced due to off-axis skewness introduced in optical fields. The resultant offsets are thus not gaussian-distributed. The maximum offsets are for sources detected at large off-axis angles on ACIS chip 2 [I2], chip 3 [I3] and chip 6 [S2]. The areal source density at the medium-deep fluxes of our sample is low enough that identification is unambiguous for 
most sources, and for the remaining few, we have associated counterparts based on comparison with studies of the same fields in the literature and accounting for source morphology in a number of filters. 


The probability of a false match occurring by chance was estimated by a \lq randomstep\rq\ method similar to that used by \citet{hornschemeier01}. Astrometric cross-correlation with DSS images was repeated after all X-ray sources in three fields were offset 10 arcsec to the north-east, north-west, south-west and south-east. The number of false source matches, averaged over the four offsets, was small: $\le 0.5$ (while the number of source matches without any offset ranged between 10 and 20 in all fields). 


\section{Redshift Determination}
\label{sec:zphot}

\subsection{Photometric Redshifts}
Our motivation for using photometric redshifts (\zp) has grown from and has been described in our previous works (C01, C02 and G02). We have demonstrated the difficulty of determining redshifts from near-infrared spectroscopy for many hard sources, while other authors have faced similar difficulties at deeper flux levels in the optical \citep[e.g., ][]{hasinger03}. Arguably, since the X-ray sources we observe are brighter than those found in deep exposures, the spectra of their counterparts should also be easier to study. This is certainly true, and the fraction of AGN for which we were able to determine redshifts from optical spectroscopy was much higher (12 of 13 in the field of A\,2390; see C02) than that of the deep fields. However, this did require the use of the Keck Telescope, and the strengths of the emission lines in many cases were small enough not to have been detected with a 4-m class telescope. We also note that while \citet{willott03} were able to find emission lines for two powerful, obscured AGN in the ELAIS field, the source classified by them as a type 2 quasar (N2\_25) was inferred to have a dust:gas ratio 10 times less than the typical value found by \citet{maiolino01} for Seyfert 2s. Thus, it is not surprising that optical AGN emission lines be visible for such a source. Instead, the one radio-quiet, resolved source (N2\_28) for which \citeauthor{willott03} did not find any line emission is more akin to the sources that we select. 

The basic \zp-determination technique has also been described in detail in C02. We use the publicly-available code HYPERZ \citep{hyperz}, which is based on fitting template SEDs to broad-band photometric fluxes in as many filters as possible. 
The template SEDs that we pass to the code include synthetic \citep{bruzual93} and empirical \citep{CWW} galaxy spectra, as well as a composite template quasar spectrum \citep{francis91}. Confidence of redshift determination depends upon: 1) photometric errors, 2) the number of filters used and 3) the representativeness of the template SEDs with respect to the sources being studied. \citet{gonzalez02} have studied the effectiveness of the photometric redshift technique applied to X-ray luminous sources, and found that while quasar-dominated sources are easy to discriminate, photometric redshift estimates agree well with spectroscopic measurements for objects in which galactic light dominates the optical flux. These latter objects are the ones that correspond to hard and optically-dim sources in our sample. Our \zp\ solutions are listed in Table~\ref{tab:zphot} on p.~\pageref{tab:zphot}ff and are plotted for each source alongside the thumbnail images (Appendix A of the electronic version).



The main advantages of including NIR photometry are that spectral breaks such as the 4000\AA\ break associated with evolved stellar populations can be properly encompassed at $z\gtsim 1$ and the shape of the continuum better constrained (see also \citealt{fernandez-soto99}). Six filters (typically $BRIJHK$, but there are variations) have been used to obtain a photometric redshift for 28 sources and five filters for 20 sources. The worst case is 1 source with deep detections/limits in three bands and very shallow limits in two more, while the best cases (3 sources) use detections/limits in seven filters (including $U$ in 2 cases and the 6.7\micron\ $ISOCAM$ LW2 filter in 1 case). Confidence of good redshift determination increases for sources with galaxy-like SEDs (i.e. harder sources), and at face value, we consider that at least the majority of very hard sources have reliable photometric redshifts, especially those with significant ($\sim$10$\sigma$) photometric detections or relatively deep limits in several bands -- these number approximately 20.

\subsection{Comparison with spectroscopic redshifts}

\begin{figure*}
  \begin{center}
    \includegraphics[angle=90,width=12cm]{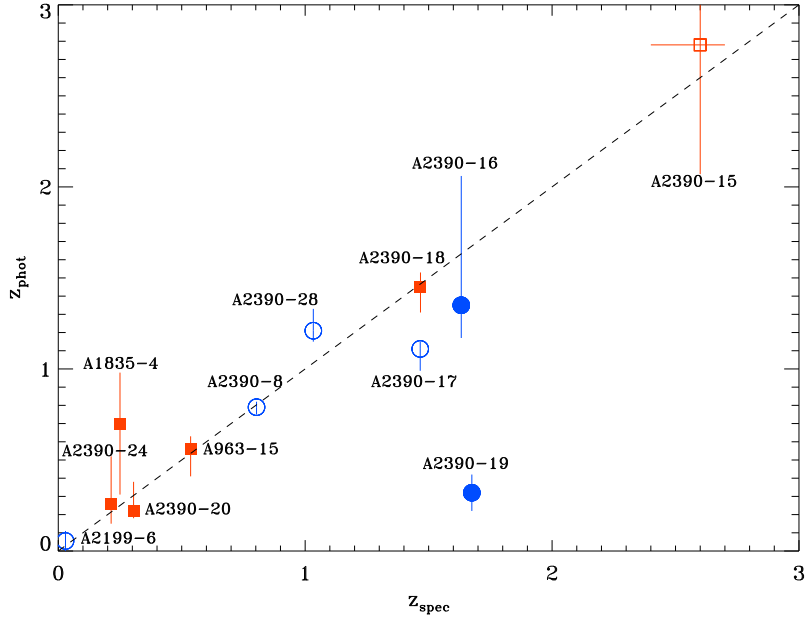}
    \caption[\zphot\ vs. \zspec]{\label{fig:compare} Comparison of photometric redshift estimates with spectroscopic measurements where both are available. Red squares are sources that possess hard or very hard X-ray counts while soft X-ray sources are represented as blue circles. The filled circles are ones which additionally possess broad optical emission lines. The unfilled red square at \zphot=2.78 is source A2390\_15, for which we do not possess a spectroscopic redshift, but have compared our redshift estimate with the estimate of \citet[][discussed in \S~\ref{sec:otherqso2}]{cowie01}. Error bars represent 90 per cent confidence intervals of the \zphot\ estimates. Table~\ref{tab:zphot} lists details of the \zp\ solutions.}
  \end{center}
\end{figure*}

We have \zspec\ measurements of 11 sources with \zphot\ estimates (see Fig~\ref{fig:compare} and Table~\ref{tab:zphot}). These are A2390\_8, \_16, \_17, \_18, \_19, \_20, \_24 and \_28 in the field of A\,2390 (C02), a bright $z=0.02929$ galaxy identified with MCG $+07-34-048$ in the field of A\,2199 \citep{mcg07-34-048}, the bright galaxy A1835\_4 \citep{piconcelli02} and one source with a tentative spectroscopic identification in the field of A\,963 (Source 15; \zs\ measurement from \citealt{lavery93}). This last source is discussed further in \S~\ref{sec:qso2} with reference to type 2 quasars. The photometric redshift estimate agrees well with the spectroscopic
measurement
for the four sources in the field of A\,2390 with narrow-line optical spectra identified by C02 (A2390\_8, \_18, \_20 and \_24) and fairly well for one of the X-ray soft, broad-line objects, A2390\_16, but with a large degeneracy. It also agrees well for the hard source A963\_15. 
The redshift estimates using galaxy templates do not agree so well with the
spectroscopic redshifts of soft X-ray sources A2390\_17 and A2390\_28, and the estimate is particularly bad for A2390\_19; in all three cases the secondary solution to the photometric redshift is no better. 

Essentially, we find that for 4 of the 5 hard X-ray sources, there is good redshift agreement (Fig~\ref{fig:compare}). This is consistent with the assumption that the obscuring gas and associated dust absorbs and scatters AGN light out of our line-of-sight. Thus, the optical-NIR flux from these sources are likely to be dominated by the host galaxy itself, and can be fit well with template galaxy SEDs.

Of course, there need not be uniform correspondence between the X-ray absorption and optical reddening -- A2390\_19 and A2390\_8 are two such sources (see C02 for details). The extinction \av\ expected from the X-ray obscuration measured for A2390\_19 is more than 10 times higher than estimated from the (very blue) optical colours used for the \zp\ determination. This source is then typical of those found by \citet{maiolino01} with a dust:gas ratio smaller than the galactic value. On the other hand, A2390\_8 -- a soft source with good redshift agreement -- possesses weak and narrow emission lines and may be example of a source with a relatively high dust:gas ratio (\citealt{barcons02} also found an AGN with similar qualitative properties, requiring a complex distribution of gas and dust).

A1835\_4 is the only hard X-ray source for which the two redshift values differ at more than the 90 per cent level. We identify this with source 41 of \citet{piconcelli02} -- even though there is a 3 arcsec offset between our coordinates, this is by far the most likely bright counterpart in the field. These authors find an elliptical morphology for this source, but the spectrum is not published as yet. Our large broad-band optical colour $B-R=2.4$ determines the estimate \zp=0.7, as compared to the measured \zs$\approx$0.25. One possible explanation for the disagreement is that our $B$ and $R$ magnitudes are derived from weak DSS constraints only, since this source lies far (11~arcmin) from the cluster core of Abell~1835 and no other archival images were available. Comparison with the spectrum when published could locate the discrepancy.

Three soft X-ray sources with disagreement show broad Mg II \l2798 (and other) optical emission lines: 1) A2390\_28, where the mismatch is only 17 per cent; 2) A2390\_16, which agrees at 90 per cent, but with a large confidence interval; 3) A2390\_19, a type 1 AGN with a dust:gas ratio different from the Galactic value. The final soft source, A2390\_17, has a complex optical morphology (see C02).
Thus, if there is evidence from optical and/or X-ray observations that AGN emission dominates the broad-band magnitudes, agreement between the measured and estimated redshifts is not necessarily expected (see also \citealt{thesis} for a more detailed discussion).


\section{Results}


Our previous work (\citealt{c02} [C02] and \citealt{g02} [G02] in particular) has shown that selection of dim optical sources is biased against any objects which lie at (or closer than) the redshifts of the clusters themselves. While the \zp-distribution is peaked at $z\sim 1$ (\S~\ref{sec:zdist}), the clusters themselves lie around $z\sim 0.05-0.3$. Of the 18 non-stellar sources for which we have been able to derive redshifts from optical or NIR {\em spectra} (of which 4 are tentative identifications and 14 are definitive), only 5 lie at $z\ltsim 0.3$, and all 5 of these have optical identifications on the DSS at least 1 magnitude brighter than the flux limit, i.e. these particular ones do not fall in the category of \lq dim\rq\ sources (two of these happened to lie in the 90 arcsec UFTI field of view while imaging other primary sources). Thus, we are able to preferentially select background AGN. This is in contrast to the cluster AGN fraction which emerges at brighter fluxes: \citet{martini02} found that 6 of 8 AGN in the field of Abell~2104 lay in the cluster, but all with $R<20$.

\begin{figure*}
  \begin{center}
\includegraphics[angle=90,width=12cm]{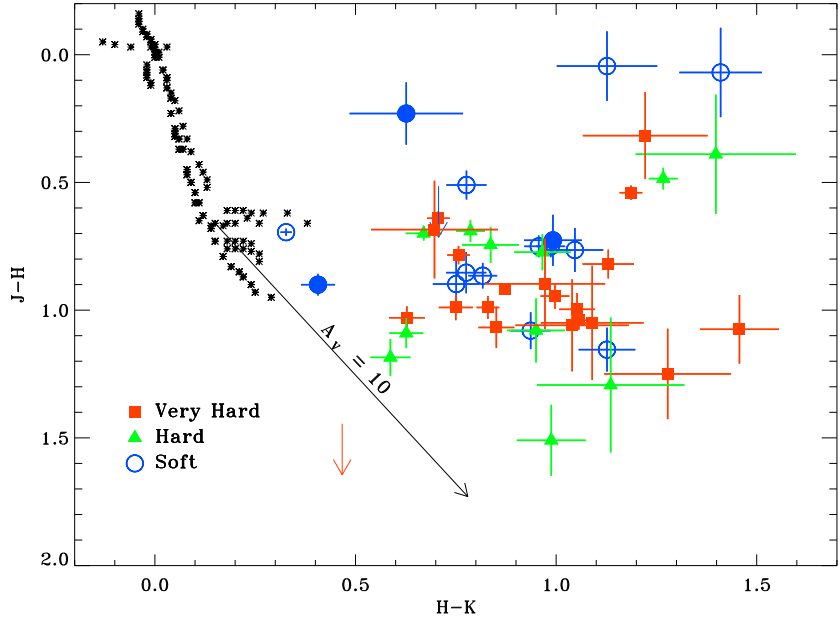}
  \caption[$J-H$ vs $H-K$]{$J-H$ vs $H-K$ for all sources with detections in at least two of the NIR bands. Colours indicate X-ray S/H ratios as follows -- red filled squares: 0$<$S/H$<$1.5 (very hard); green filled triangles: 1.5$<$S/H$<$2.5 (hard); blue open and filled circles: S/H$>$2.5 (soft). The three filled circles are soft sources spectroscopically identified as type 1 broad-line AGN: A2390\_16, \_19 and A1835\_2. The asterisks represent colours of main sequence, giant and super-giant stars of all spectral types from O (with the most blue colours) to M (with the reddest colours). The stellar colours are from \citet{allen}, tabulated at http://www.jach.hawaii.edu/JACpublic/UKIRT/astronomy/. The arrow shows the sense in which extinction with \av=10, assuming a Galactic extinction law, would change the colours. }
  \label{fig:jh_hk}
\end{center}
\end{figure*}

Our sample is strongly biased against detection of stars. While \citet{c02} found two X-ray sources with stellar optical spectra (out of 15), both of these possessed soft X-ray colours. The combination of selection based on hard X-ray count ratios and optically-dim magnitudes is likely to filter out most stars. Additionally, as Fig~\ref{fig:jh_hk} shows, the NIR colours of our sample are different from those of unreddened stars. The sources closest to the stellar locus are soft X-ray sources with broad optical lines and/or blue broad-band continua similar to quasars.
Moreover, only 11 of the 56 sources with \k-band detections presented here have unresolved (stellar) morphologies consistent with the seeing measurements on the night of observation, and the spectra of 5 of these show that they are a mixture of AGN and galaxies. 

\subsection{Colours and Fluxes}

\begin{figure*}
  \begin{center}
\includegraphics[angle=90,width=12cm]{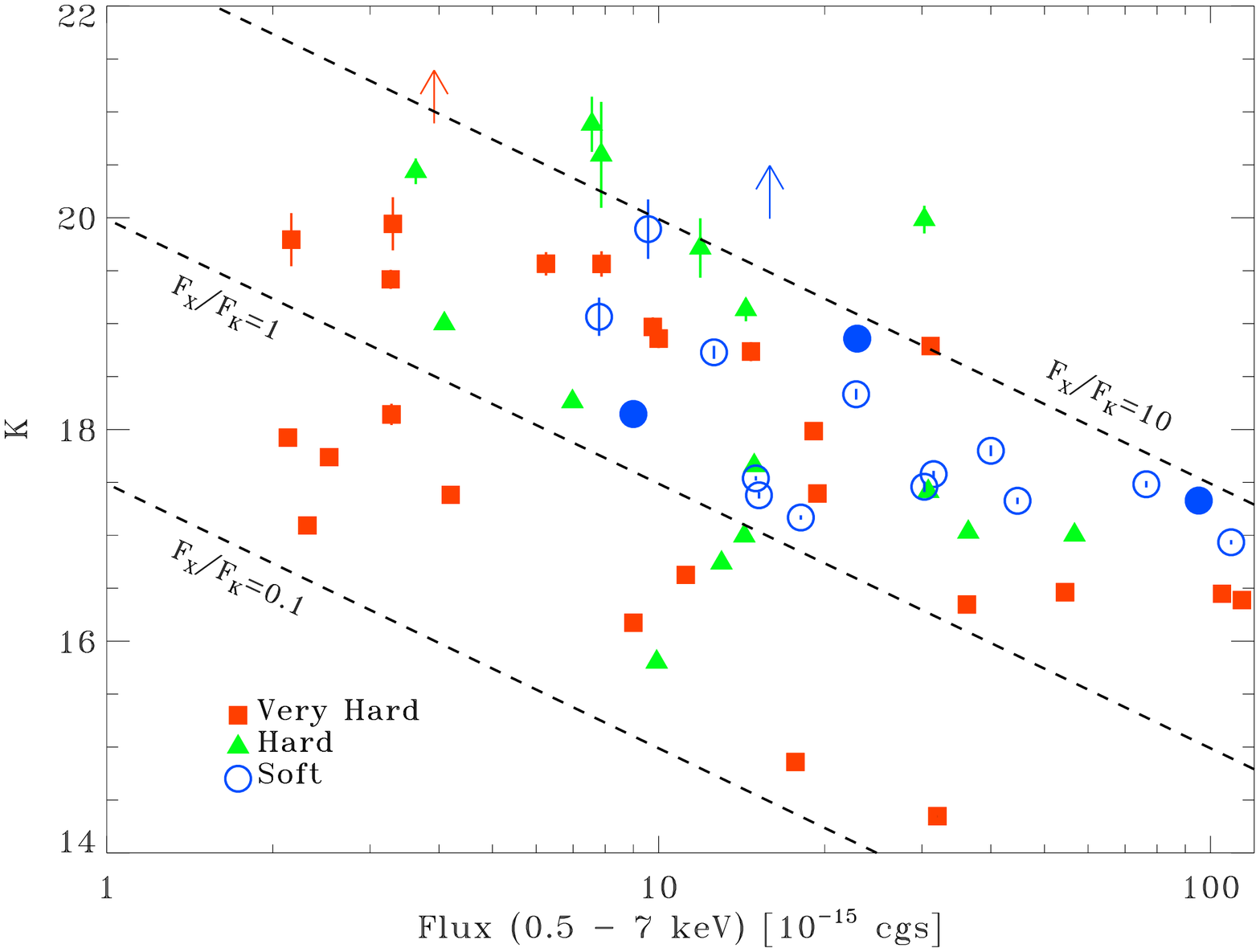}
  \caption{\k-magnitude vs X-ray flux for 58 imaged X-ray sources. The symbols are as in the previous figure. The lowest redshift source in our sample, identified with MCG$+$07$-$34$-$048 at $z=0.02929$ is off the scale at [4.3, 12.1]. 
}
  \label{fig:k_fx}
\end{center}
\end{figure*}

Fig~\ref{fig:k_fx} shows the observed \k-magnitudes of the counterparts to 58 X-ray sources followed-up (including limits for 2 non-detections in the field of A\,963) against their 0.5--7 keV band X-ray fluxes\footnote{To aid comparison with published results in other bands, we list the following flux ratios for the photon-indices given in brackets: $F_{2-10}:F_{0.5-7}=$ 0.96 (1.4), 0.72 (1.8), 0.61 (2.0); $F_{2-8}:F_{0.5-7}=$ 0.77 (1.4), 0.61 (1.8), 0.53 (2.0).}. The fluxes are computed by assuming a $\Gamma=1.4$ power-law under Galactic absorption only. Two sources for which this assumption leads to a significant underestimate of the flux compared to the values measured from their extracted X-ray spectra are A2390\_18 and A963\_15 (the true flux is higher by factors of 1.8 and 3.1 respectively; \S~\ref{sec:qso2}), indicating that they are better described by harder X-ray power-laws. Thus, for these two sources we use the directly measured fluxes.
The figure shows that all sources bracket the break in the source counts at 10$^{-14}$ \ergpspsqcm\ within \p1 dex. The large X-ray fluxes relative to the \k-band flux are consistent with powerful accretion activity. For less-energetic non-accretion activity, \citet{alexander_red} found an X-ray:optical flux ratio$\approx$0.01 (see also \citealt{hornschemeier03}) -- the X-ray:NIR is likely to be similarly small for optically-dim sources without ongoing accretion. Hard sources (squares and triangles in Fig~\ref{fig:k_fx}) have a larger spread with a lower average X-ray:\k\ ratio than the soft sources (blue circles), and are consistent with obscured AGN in which the NIR flux is dominated by the host galaxy, while the X-rays are scattered out of the line-of-sight by the obscuring matter itself.


\begin{figure*}
  \begin{center}
\includegraphics[angle=90,width=12cm]{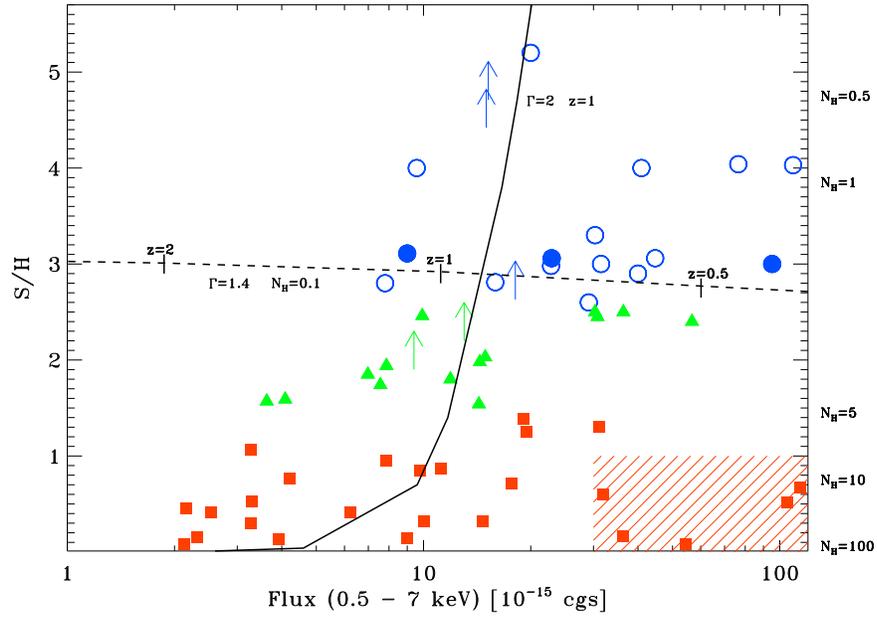}
  \caption[X-ray S/H ratio vs X-ray flux]{X-ray S/H ratio vs X-ray flux for sources in Fig~\ref{fig:k_fx}, with the same symbols. The dashed line is the S/H ratio predicted for a power-law source with $\Gamma=1.4$ and moderate ($0.1\times 10^{22}$ cm$^{-2}$) intrinsic absorption with decreasing flux, i.e. placing it at increasing redshift. Three redshifts are marked. The solid line is a $\Gamma$=2 source with an intrinsic luminosity of 10$^{44}$ \ergps\ at a fixed $z$=1, showing the effect of increasing column density local to the source (shown on the right in units of $10^{22}$ cm$^{-2}$). Additionally, five X-ray detected sources in the field of A\,2390 with optical spectra (from which we have measured redshifts) but no NIR imaging are included. These are: A2390\_4, \_6, \_10, \_13, \_26. The hatched region on the lower right is where bright and hard sources (including type 2 quasars) emerge.
}
  \label{fig:sh_fx}
\end{center}
\end{figure*}

But there are also hard sources with bright X-ray fluxes and powerful \k-band emission toward the lower right of the figure. These are better distinguished in a plot of the S/H ratio vs. X-ray flux shown in Fig~\ref{fig:sh_fx}, where two main features are evident. Firstly, the general trend of softer sources being the brighter ones: this is due to the obscuration of the AGN in the harder sources which also diminishes the AGN brightness. This hardening of counts is illustrated by the solid line, which shows the S/H ratio predicted for an AGN with \lx (0.5--7 keV restframe)=10$^{44}$ \ergps\ at the fixed redshift $z$=1. Column density increases along the line from top to bottom (marked along the y-axis on the right in units of 10$^{22}$ cm$^{-2}$). In contrast, an unabsorbed source with a constant luminosity and observed at different redshifts will not become harder with redshift (dashed line), but may get softer due to the hard counts shifting into the soft band, if $\Gamma<2$.

The second trend to be noticed is the broader range of fluxes occupied by the hardest sources (red squares). While their fluxes are biased towards the faint end, a population of hard X-ray sources with fluxes $\ge 3 \times 10^{-14}$ \ergpspsqcm\ is also seen at the brightest end. Most of these are also hosted in bright galaxies. In fact, all five very hard (S/H$<$1) sources in this regime (hatched region of Fig~\ref{fig:sh_fx}) have \k$<$16.5. These are A2390\_18, \_20, A2204\_1, A1835\_4 and A963\_15. The first two sources lie at $z$=1.467 and 0.305, a type 2 quasar and narrow-line AGN respectively. The next two have \zphot=0.5 and \zs$\approx$0.25 respectively, and A2204\_1 has a very flat \k-band spectrum and red optical/NIR colours consistent with high obscuration (G02). The final source, A963\_15 at $z$=0.536 (\S~\ref{sec:a96315}) is also a type 2 quasar. 
Of these, only A1835\_4 has an inferred unabsorbed X-ray luminosity $L_{\rm X}<10^{44}$ erg s$^{-1}$ -- thus, at these bright and hard fluxes, powerful and obscured quasars as well as Seyferts at a range of redshifts ($z\sim 0.3-1.5$) can be easily distinguished.

We note that another hard (S/H$\approx$1.3) source in this bright regime is the other type 2 quasar A2390\_15, which, however, lies at \zphot=2.8 in a very faint galaxy and is lensed by a factor of 7.8 (\S~\ref{sec:qso2}).

\begin{figure*}
  \begin{center}
\includegraphics[angle=90,width=12cm]{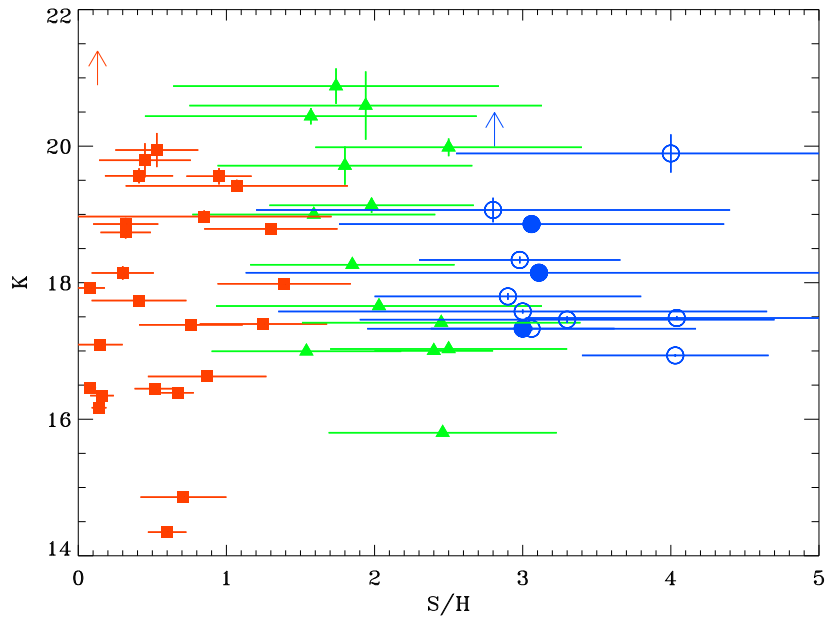}
  \caption[\k\ vs X-ray S/H ratio]{\k\ vs X-ray S/H ratio for the imaged sources, with symbols as in Fig~\ref{fig:jh_hk}. No obvious trend is observed between the two plotted quantities. The error-bars represent 1-$\sigma$ errors. To reiterate, colours indicate X-ray S/H ratios as follows -- red squares: 0$<$S/H$<$1.5 (very hard); green triangles: 1.5$<$S/H$<$2.5 (hard); blue open and filled circles: S/H$>$2.5 (soft). 
}
  \label{fig:k_sh}
\end{center}
\end{figure*}

If the NIR light from most X-ray sources, especially the hard ones, is dominated by the host galaxy itself, then no direct correlation need be present between this NIR flux and X-ray obscuration column-density. This is indeed seen to be the case in Fig~\ref{fig:k_sh}. We also take the opportunity to show in this figure the errors on the S/H ratios themselves (not shown in the previous plot for clarity), which are dominated by poisson noise on the detected counts. While these large errors make it impossible to correctly classify many sources, the boundaries of the X-ray count ratio are such that all sources with very hard X-ray colours (S/H$<$1.5; red squares) could be classified as hard (i.e. S/H$<$2.5) at the 95 per cent level. 

\subsection{Redshift Distribution}
\label{sec:zdist}

\begin{figure*}
  \begin{center}
\includegraphics[angle=0,width=12cm]{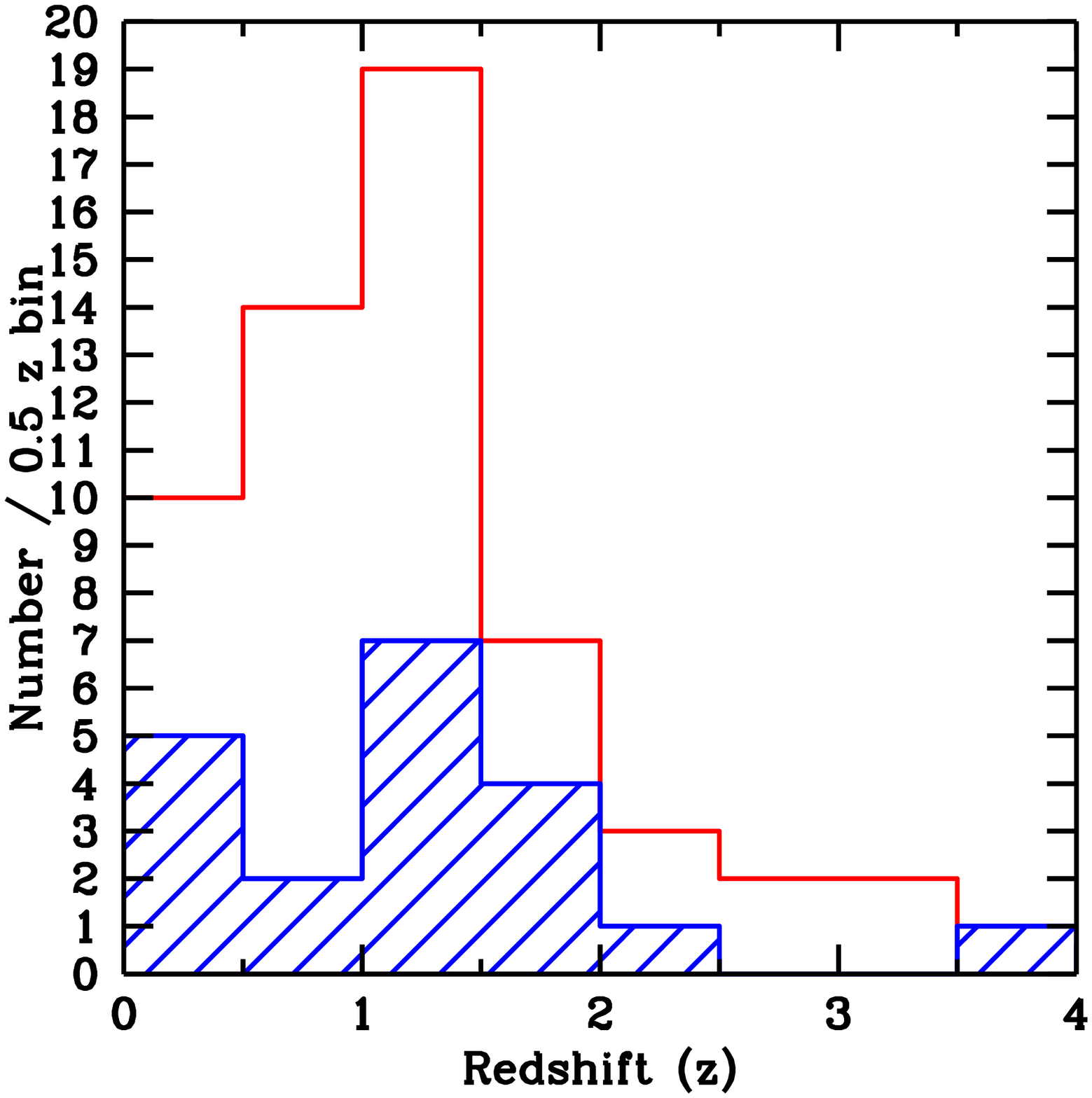}
  \caption[Redshift Histogram]{Redshift Histogram for 20 sources with spectroscopic measurements (blue hatched regions) and 38 additional sources with photometric redshifts only (red outlined region). Four sources with spectroscopic redshifts which are tentative measurements are: A963\_15 (at z=0.536) -- redshift determined from \citet{lavery93} and found consistent with the estimated \zphot; A1835\_1 (z=1.256), A1835\_2 (z=3.830) and Per\_2 (z=1.307) -- only single emission lines were detected for these three and consistency is found from \zphot\ estimates for the two in the field of Abell~1835 and detection of an X-ray Fe K$\alpha$ line for the last source (see \citealt{thesis} for more details).
}
  \label{fig:zhist}
\end{center}
\end{figure*}

The redshift distribution derived for our sample is shown in Fig~\ref{fig:zhist} where the histogram shows the spectroscopic redshift if known, otherwise the best photometric redshift. For completeness, we also include the 5 sources from the field of Abell 2390 for which we have determined a redshift from Keck spectra (C02), but do not have NIR imaging, bringing the total number of redshift identifications to 58. We find a distinct peak at $z\sim 1$. Though we do not have enough sources for a finer binning in redshift, we note that this result is in broad agreement with the findings of deep field follow-up work, where a significant fraction of \lq type 2\rq\ sources are seen to lie at redshifts lower than the peak at $z\sim $1.5--2 predicted by XRB synthesis models (by matching to the unobscured quasar distribution; \citealt{comastri95}, \citealt{mi00}).

\begin{figure*}
  \begin{center}
\includegraphics[angle=90,width=12cm]{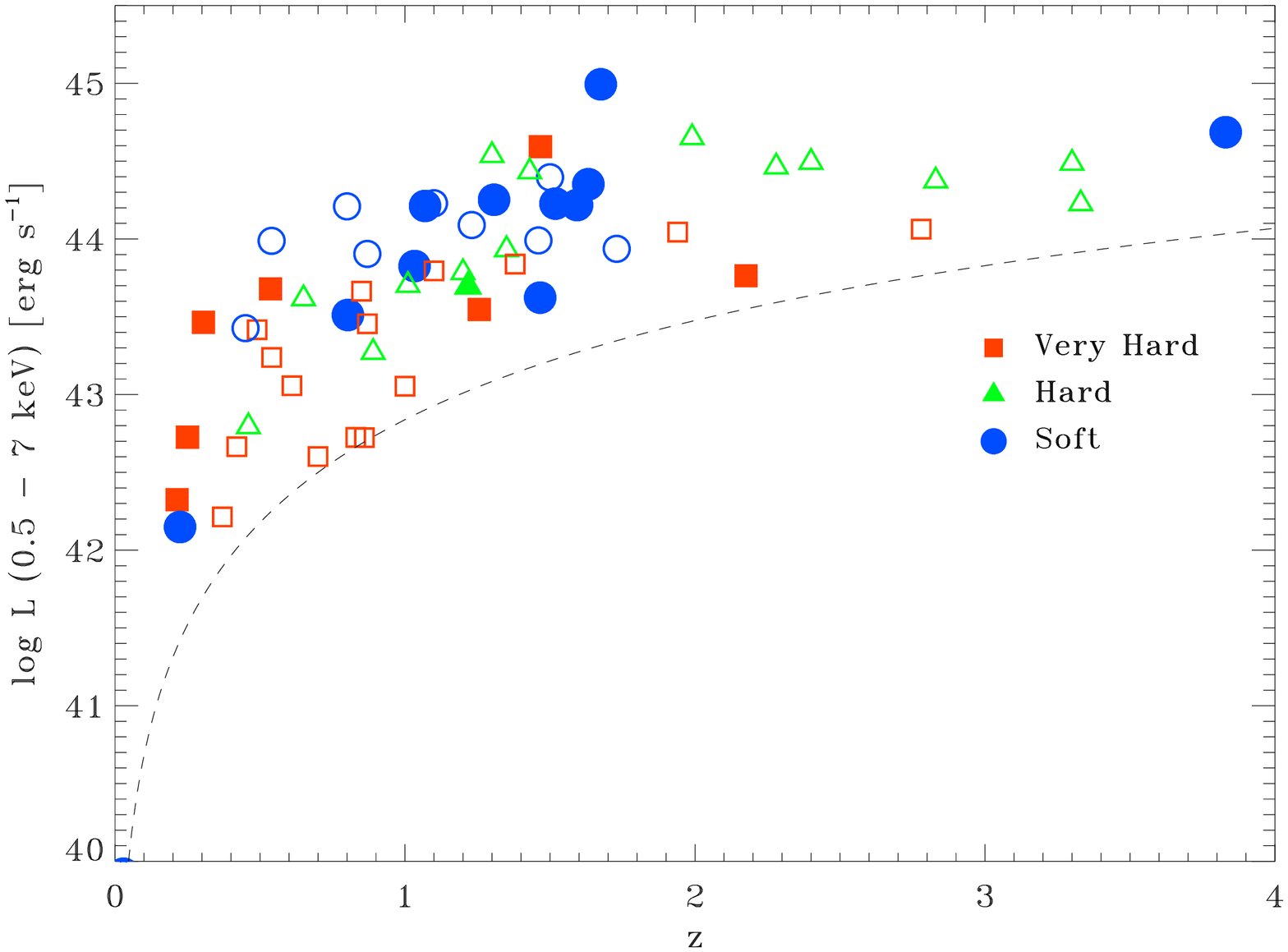}
  \caption[X-ray luminosity vs. $z$]{$L_{\rm X}$ (0.5--7 keV restframe) vs $z$ for all sources with both measurements. Luminosities have been calculated from the fluxes in Figs~\ref{fig:k_fx} and \ref{fig:sh_fx} and are de-magnified for A2390\_15, A2390\_17 and A2390\_18. No correction has been made for the local absorbing column. The filled symbols denote sources with spectroscopic redshifts, the rest being photometric estimates. The dashed line is a rough estimate of the faintest limiting luminosity of our sample, shown for a source with full band X-ray flux of 2$\times$10$^{-15}$ \ergpspsqcm, as a function of redshift.
}
  \label{fig:lx_z}
\end{center}
\end{figure*}

Shown in Fig~\ref{fig:lx_z} is the 0.5--7 keV X-ray luminosity of the sample sources versus redshift, assuming \zp\ to be the true redshift if no spectroscopic redshift is available. The luminosity is calculated from the fluxes of Figs~\ref{fig:k_fx} and \ref{fig:sh_fx} (typically assuming $\Gamma=1.4$) and de-magnification is applied to the lensed sources in the field of A\,2390. Assuming a steeper power-law (say, $\Gamma=2$), will decrease the luminosity, while correction for obscuration will increase it. While the \c\ observations can detect powerful Seyferts with $L_{\rm X}\sim 10^{44}$ erg s$^{-1}$ out to $z=3$, we are able to detect Seyferts with $L_{\rm X}\sim 5\times 10^{42}$ erg s$^{-1}$ to $z=0.8$. Note that the median luminosity of AGN in the deep \c\ fields is $\sim 10^{43}$ erg s$^{-1}$ \citep[e.g., ][]{hasinger03}.

There is no significant distinction between the distributions of sources based on their S/H ratios, except a slightly lower median redshift for the very hard sources (red squares; $z_{\rm median}=0.8$) compared to the soft sources (blue circles; $z_{\rm median}=1.2$). Beyond $z\approx 2$, five of the eight sources seen are X-ray hard (1.5$<$S/H$<$2.5). Without redshift identifications from a complete sample, we cannot comment on this distinction with certainty.

\begin{figure*}
  \begin{center}
\includegraphics[angle=90,width=12cm]{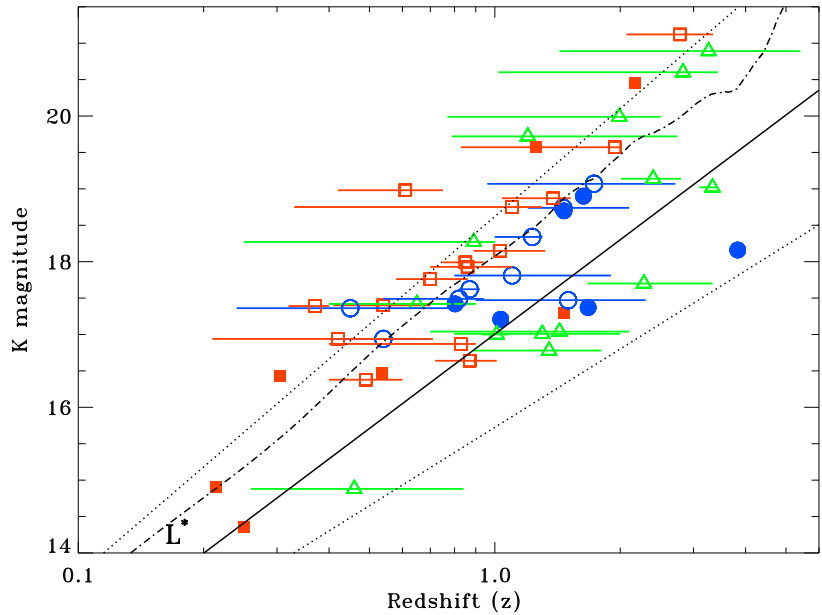}
  \caption[$K$ vs $z$ for all sources]{$K$ vs $z$ for all sources with both measurements. The
solid line shows the $K$--$z$ relation for massive radio ellipticals, and the
dotted lines indicate the scatter of the radio galaxies about this
line (taken from \citealt{eales93}). The dot-dashed line shows the predicted curve for an $L^*$ galaxy formed in a single stellar burst at $z$=10. The 90 per cent \zp\ confidence intervals are shown for sources with only a photometric redshift. As in Fig~\ref{fig:lx_z}, filled symbols denote spectroscopic redshifts. Strongly lensed sources in the field of Abell~2390 have been de-magnified.
}
  \label{fig:Kz}
\end{center}
\end{figure*}

Fig~\ref{fig:Kz} shows the $K$-magnitude versus redshift of the counterparts to the X-ray sources. Confidence intervals of the \zp\ calculation are also shown. It is clear that our sample covers the brighter region of $K$--$z$ space, and thus probably includes massive and bright galactic systems, which also host very massive black holes (c.f. \citealt{magorrian98, eales93}; or \citealt{willott03kz} for a more recent determination of the $K$--$z$ relation). Several sources are up to two mags brighter than $L^*$ 
(while at the other extreme, one source, RXJ0821\_3, is fainter than $L^*$ by two mags).

\begin{figure*}
  \begin{center}
\includegraphics[angle=90,width=12cm]{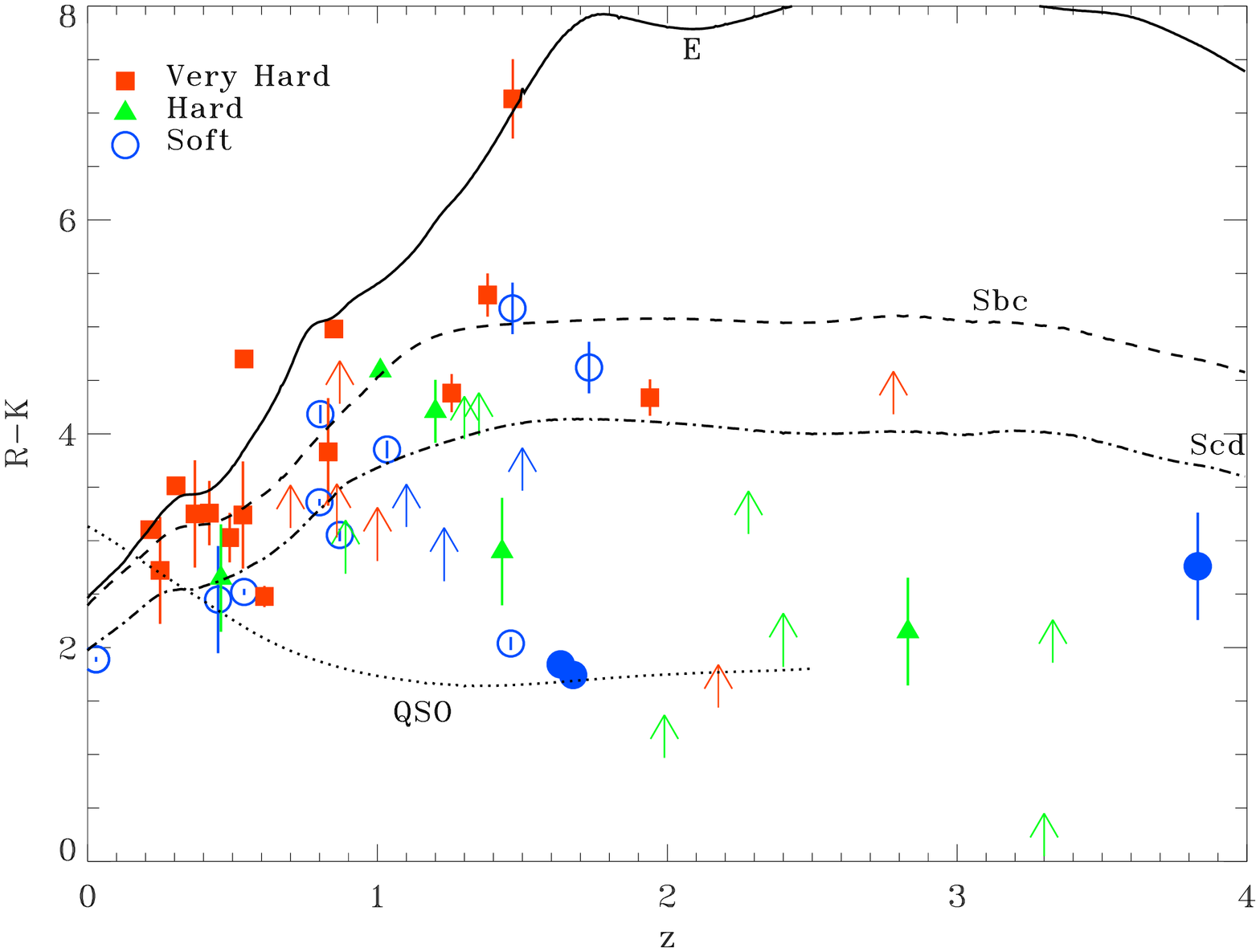}
  \caption[$R-K$ vs z for all sources]{$R-K$ vs z for all sources with $K$ band detections and $R$ band detections or limits, and a redshift determination. All colours have been
  corrected for Galactic line-of-sight reddening. Colour tracks are
  shown for unreddened Coleman-Wu-Weedman (CWW) E (solid), Sbc
  (dashed) and Scd (dot-dashed) empirical templates. The dotted line
  is a colour track of a radio-quiet quasar from \citet{elvis94}. The three filled circles are broad-line AGN.
}
  \label{fig:rk_z}
\end{center}
\end{figure*}

Additional evidence that the optical/NIR light in many sources is dominated by the host galaxy comes from a plot of the optical--NIR colour vs. $z$. Although many $R$ magnitudes in our sample have weak limits only, we present the $R-K$ colour to facilitate comparison with the many publications which plot this colour (we also have fewer $I$-bands measurements compared to $R$). Fig~\ref{fig:rk_z} shows that, while there is some degeneracy of the data between the various models, the vast majority of hard sources have colours which are consistent with those of galaxies rather than those of a quasar (or limits consistent with this trend). In contrast, the three sources in which broad emission lines have been identified all possess very blue colours. 
As mentioned earlier, X-ray hardness need not correlate directly with optical reddening. Such a mismatch could indicate either a dust:gas ratio for the AGN obscuring matter different from the Galactic value, or contamination by other (non-accretion) sources of radiation, e.g., starburst activity on larger scales. Such sources may be located with plots such as Fig~\ref{fig:rk_z}, but a good measure of optical reddening will require measurement of emission line ratios, such as the Balmer decrement. We note that there are only 4 sources with $R-K\gtsim5$ (extremely red objects; the reddest one is the type 2 quasar A2390\_18), but the many shallow colour limits may be hiding more such sources [the deep \subaru\ photometry of \citet{cowie01} shows that A2390\_15 (their source 3) is also extremely red, while our limits are only $R-K>4.4$].

\subsection{Type 2 quasars}
\label{sec:qso2}

Having discussed some general properties of the sample, we now focus on the most luminous of the hard sources. 


\subsubsection{A2390\_18: \xmm\ PN exposure}

\begin{figure*}
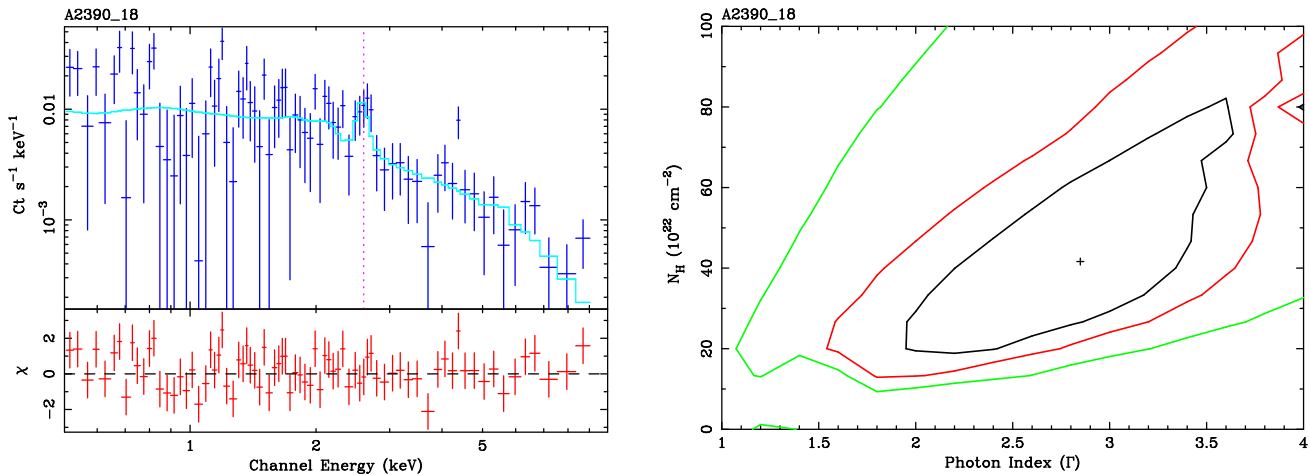

  \begin{center}
    \includegraphics[angle=270,width=8cm]{a18pn_mekalzpo_goodaxes.ps}
\hspace*{0.5cm}
    \includegraphics[angle=270,width=8.6cm]{a18pn_mekalzpo_contours.ps}
    \caption[\xmm\ PN spectrum of A2390\_18]{(Left) \xmm\ PN spectrum for A2390\_18. An 8 keV MEKAL model fixed at $z=0.228$ accounts for emission below 1 keV. The source is modelled with an absorbed power-law, intrinsic obscuration and an Fe K$\alpha$ line at $z=1.467$, whose expected redshifted energy is marked with the pink, dotted line.\\
(Right) 68, 95 and 99 per cent confidence contours for the redshifted absorbed power-law model. The evidence for intrinsic obscuration above the Galactic column is very significant.} \label{fig:a18xmmpn}
  \end{center}
\end{figure*}

One clear type 2 quasar was presented in \citet{f00} and elaborated upon in C02 (source A2390\_18 with $N_{\rm H}=2\times 10^{23}$ cm$^{-2}$ and intrinsic $L_{2-10}\approx 10^{45}$ \ergps, assuming the standard non-zero Lambda cosmology; see Fig~12 of C02 for the \c\ ACIS spectrum). This is also the reddest source in our sample, with $R-K>7$ (Fig~\ref{fig:rk_z}). Recently, we have also obtained \xmm\ data on this field (effective exposure time 18 ks), and extracted a spectrum of this source from the PN chip (Fig~\ref{fig:a18xmmpn}). Background flare removal, pattern selection and spectral extraction were all performed using the XMM-Newton Science Analysis Software\footnote{http://xmm.vilspa.esa.es/external/xmm\_sw\_cal/sas.shtml} and the standard on-axis response matrix. A power-law transmission model at $z=1.467$ fitted to the data implies a hard spectrum, with a free-fitting $\Gamma=1.2$ with no absorption required above the Galactic column. The XSPEC package \citep{xspec} was used for all fitting of X-ray spectra. The data were binned with a minimum of 15 counts per bin, and the $\chi^2$/dof\footnote{dof $\equiv$ degrees of freedom} goodness-of-fit is 
80/74. 
The spectrum shows, however, that there is a large amount of soft emission below 1 keV, which is not expected in a highly absorbed source (compare with the \c\ spectrum). This excess is due to incomplete subtraction of thermal emission from the cluster Abell~2390 itself -- the relatively large PSF of \xmm\ makes it difficult to accurately measure the background close ($\sim 1$ arcmin) to a bright extended source. We have accounted for this excess by incorporating a MEKAL model with a temperature of 8 keV at $z=0.228$, the redshift of the cluster A\,2390. A free-fit now gives 
$\Gamma=2.85$ [1.2, 4.1] 
but a large 
$N_{\rm H}=4.1 [0.8, 11.1] \times 10^{23}$ cm$^{-2}$ 
[Numbers in square brackets denote 90 per cent confidence intervals]. The free-fitting photon-index value is steep, but consistent with that found in the ACIS fit ($\Gamma=3.2\ [2.7, 4.5]$; C02). With $\Gamma=2$ fixed, the absorption implied is 2.2 [1.1, 4.0] $\times 10^{23}$ cm$^{-2}$. In addition, there is an excess of counts close to the energy where a redshifted 6.4 keV (rest-frame) neutral Fe K$\alpha$ line is expected to lie. The line is significant at more than 99 per cent according to an f-test, but the preferred rest-energy is 6.48 [6.3, 6.6] keV. A simple narrow redshifted gaussian line fit to this excess implies a rest-frame equivalent-width $\sim 1$ keV. Such strong lines are typically indicative of a reflection-dominated spectrum. On the other hand, letting the width of the line vary implies that the line is broad: $\sigma=0.12$ keV, but a low f-test probability for this suggests that we cannot contrain the width with the current data. A longer observation with more target photons may resolve the Fe line further. We note that the non-detection of any obvious Fe line in the \c\ spectrum (Fig~12 of C02) is consistent with the fewer counts observed ($\approx 70$ source counts in each of two ACIS exposures). 

\subsubsection{A963\_15}
\label{sec:a96315}

\begin{figure*}
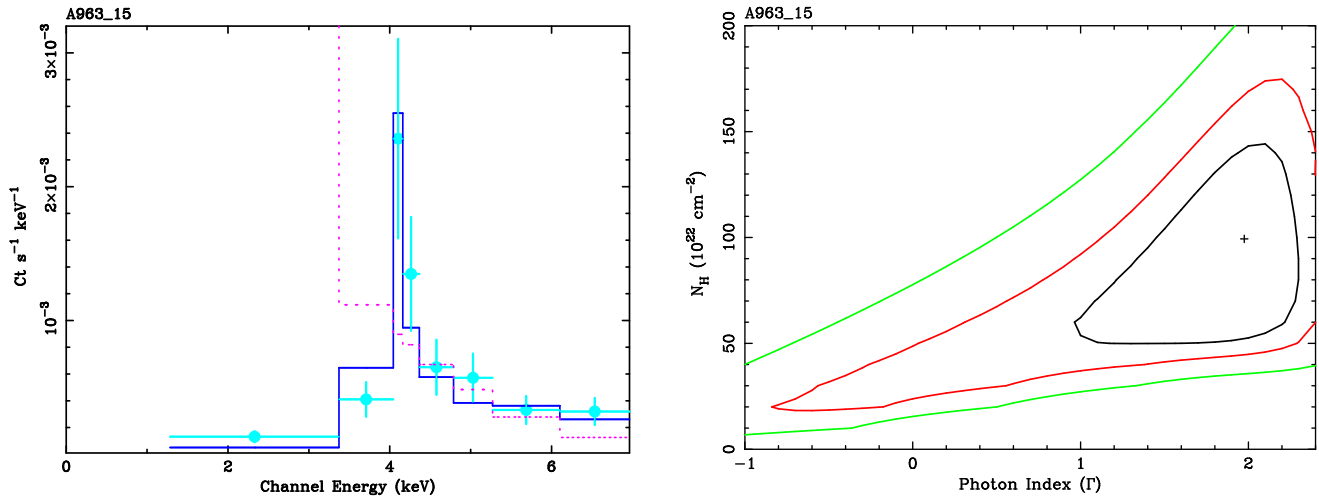

  \begin{center}
\includegraphics[height=8.3cm,angle=270]{a963_src15_z05wapogar_good.ps}
\hspace*{0.5cm}
\includegraphics[width=6.5cm,angle=270]{a963_src15_z05wapogar_free_contours.ps}
  \caption[A963\_15 ACIS spectrum]{(Top) \c\ ACIS-S3 spectrum of source 15 in the field of Abell 963. This is a very hard source with S/H=0.08 and 90 counts in the 0.5--7 keV band. The data (blue circles with 1$\sigma$ errors) has been fit to a power-law transmission model (dark blue line) with $\Gamma=2$ and a 6.4 keV Fe K$\alpha$ line at $z=0.536$. The fitted intrinsic absorption \nh$=1.1\times 10^{24}$ cm$^{-2}$. The pink dotted line is the power-law model affected only by Galactic line-of-sight obscuration. The flux of spectrum of A963\_15 is plotted on a linear scale to emphasize the potential emission line.\\
(Bottom) 68, 95 and 99 per cent confidence intervals for a free-fit of the intrinsic power-law photon-index and the obscuring column density fixed at $z=0.536$.
}
  \label{fig:a963_src15_xray}
\end{center}
\end{figure*}

Source 15 in the field of Abell~963, whose X-ray spectrum is shown in Fig~\ref{fig:a963_src15_xray}, is a very hard source with S/H=0.08\p0.04. Based on $UBRiJHK$ photometry, we estimate \zphot=0.56 [0.4, 0.7], which is in agreement with a tentative redshift of 0.536 found by \citet{lavery93} based on detections of recombination lines of Oxygen. Even if the redshift is fixed at $z$=0, the absorption implied is high: \lognh$>$23. Fig~\ref{fig:a963_src15_xray} shows a $\Gamma$=2 absorbed power-law fit to the data assuming the above spectroscopic redshift. The peak at 4.1 keV prompted us to include a narrow (10 eV fixed width), redshifted Fe K$\alpha$ line, and this was found to match the peak energy well, further supporting the redshift estimate. While the rest-frame equivalent width of the line is close to 1 keV, its significance is marginal (f-test gives a null hypothesis probability of 0.1). With the above fixed value of $\Gamma$=2, the X-ray absorption implied is $N_{\rm H}=1.0 [0.5, 1.5] \times 10^{24}$ cm$^{-2}$, making this source nearly Compton-thick. The implied absorption-corrected 2--10 keV luminosity is $6\times 10^{44}$ erg s$^{-1}$, making this a clear type 2 quasar detected in X-rays. Note that this fit was performed with a minimum of 10 counts per bin since only about 90 counts were detected in all ($\chi^2$/dof=7/5). Despite small-number statistics, the bottom part of Fig~\ref{fig:a963_src15_xray} shows that absorption above the Galactic column is significant -- the contours are shown for a freely-varying photon-index vs. absorption. 

On the other hand, the spectrum can be fit by a PEXRAV reflection model without any transmission. We assume a uniform geometry, in which case the reflector subtends 2$\pi$ steradians at the illumination source and the source is Compton-thick. An additional foreground absorber with a column of $N_{\rm H}=5\times 10^{23}$ cm$^{-2}$ is required and the implied intrinsic luminosity is then $1.4\times 10^{44}f^{-1}$ erg s$^{-1}$. This depends on the albedo $f$, which can be as low as just a few per cent, implying a luminosity as large as $\sim 10^{46}$ erg s$^{-1}$. We cannot distinguish between the transmission and reflection models with only about 90 counts obtained in 36 ks. \xmm, with its higher collecting area, has the potential to provide better statistics and reveal any Fe line. Unfortunately, in the 25~ks observation of the cluster Abell~963 available in the archive\footnote{http://xmm.vilspa.esa.es/external/xmm\_data\_acc/xsa/} (PI: Kneib), this source lies exactly in a chip gap on the PN camera (which has better sensitivity to hard counts). A first look at the pipeline processed files delivered with the archival products reveals $\sim$70 background-subtracted counts in each of the MOS cameras. We do not perform an extraction from the raw observation data files here.
Service observing time on the William Herschel Telescope has also been allocated to us in order to obtain an optical spectrum of this source. The study of any optical emission lines or 4000\AA\ break present would conclusively identify the redshift, and comparison of optical reddening with the X-ray absorption will provide further clues to the nature of this source.

\begin{figure*}
  \begin{center}
\includegraphics[angle=0,width=6cm]{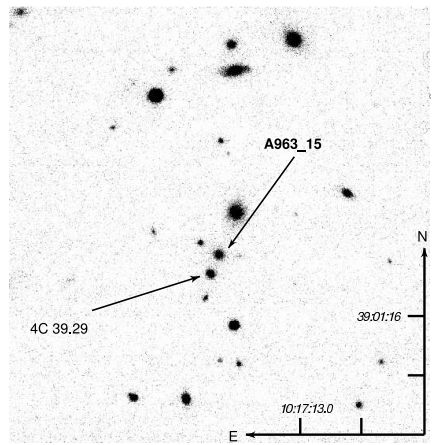}
  \caption[A963\_15: \k-band field of a type 2 quasar]{
UKIRT UFTI \k-band image of the field of source 15 in the field of Abell 963. North is up and east is to the left, and intervals of 10 seconds of arc are tickmarked along the mini-axes. The position of the counterpart associated to the radio source 4C~39.29 is also marked. Since identification in the literature is ambiguous, we present a large field-of-view $\approx$1.1 arcmin square; the J2000 coordinates of the counterpart to source A963\_15 are RA 10:17:14.1 DEC 39:01:25. 
}
  \label{fig:a963_src15_K}
\end{center}
\end{figure*}

Radio observations have detected a source (4C~39.29; \citealt{riley75}) with coordinates (as measured in the radio maps) offset from the X-ray position by 3 arcsec. This region contains an overdensity of sources (with 5 NIR neighbours brighter than K$\approx$19.5 within 8 arcsec of the X-ray position) and unambiguous identification of the radio counterpart is difficult. Comparison with \citet{lavery93}, however, suggests the configuration shown in Fig~\ref{fig:a963_src15_K}, where the X-ray source is associated with a galaxy in the centre of the field, while the radio source lies approximately 4 arcsec to the south-east.

\subsubsection{Other powerful obscured AGN in the sample}
\label{sec:otherqso2}

An additional type 2 quasar -- A2390\_15, also strongly lensed by the cluster potential -- was published by C02. Determining a spectroscopic redshift for this source with both the Keck in the optical and SUBARU in the near-infrared has proven difficult (see also \citealt{cowie01}). Photometric-redshift determination suggests $z=2.6-2.8$ (but see also \citealt{lemonon98}). C02 presented the broad-band SED for this source, covering the X-ray, optical, NIR, MIR and sub-mm regimes. Based on radiative-transfer modeling through dust using the publicly-available code DUSTY \citep{dusty}, the presence of an absorbed quasar with optical:X-ray spectral index $\alpha_{\rm OX}=1.3$ and intrinsic, lensing-corrected luminosity of $\sim 2\times 10^{45}$ erg s$^{-1}$ was inferred over the ultraviolet regime encompassing the big-blue bump \citep[assuming the standard cosmology; see also][]{wfg}.

With regard to the rest of our sample, while we do not have enough counts to extract X-ray spectra for many objects, we note that there are 6 other sources with S/H$<$2.5 (i.e. 6 green triangles in Fig~\ref{fig:lx_z}) and $L_{0.5-7}> 3\times 10^{44}$ erg s$^{-1}$, or 9 other sources above a limit of $10^{44}$ erg s$^{-1}$. The S/H ratios are consistent with spectra harder than a $\Gamma=1.4$ power-law affected by at least modest ($\sim 10^{21}$~cm$^{-2}$) absorption at all redshifts. These sources are tabulated in Table~\ref{tab:qso2} and are primary targets for further study aimed at finding type 2 quasars. Including the three sources discussed in more detail above, the total number of type 2 quasars and candidates comes to 12. 

Of course, by selecting on the basis of S/H ratios, we are biased towards the most absorbed and/or low-redshift AGN. For instance, an AGN with $\Gamma=1.4$ absorbed by a column of \nh=$10^{21}$ cm$^{-2}$ at $z=3$ has a S/H ratio of 3.0, if observed on ACIS. Similarly, an AGN with a more typical photon-index $\Gamma=2$ and absorbed by a column of \nh=$10^{23}$ cm$^{-2}$ at $z=3$ also has S/H=3.0, too soft to be included in our hard sample. 


In summary, we have been able to immediately identify and investigate in some detail a few powerful sources which are highly obscured or are at least consistent with large obscuring columns. The sample as a whole presents examples of hard sources at fluxes faint enough that they would not have been detected in previous X-ray missions, but which are important individual contributors to the X-ray background. If all such sources possess extremely weak emission features, this has important consequences for the standard AGN unification model.

\section{Discussion}

Our basic selection strategy has been to focus on the X-ray count ratios formed from a hard (2--7 keV) and a soft (0.5--2 keV) band, followed by selection based on weakness of optical flux. The hardness ratio S/H=2.5 will allow selection of obscured AGN in the following [\nh, $z_{\rm max}$] space, if they lie above our flux threshold: [$10^{22}$, 0.5], [$5\times 10^{22}$, 2.0], [$10^{23}$, 3.0] 
-- i.e., sources with an obscuring column \nh\ (cm$^{-2}$) will be selected out to the approximate redshift $z_{\rm max}$, assuming an intrinsic photon-index $\Gamma=2$. Our limiting 0.5--7 keV rest-frame luminosities corresponding to the absorbing columns and maximum redshifts above are approximately $1.5 [3]\times 10^{42}$, $2 [8]\times 10^{43}$ and $0.4 [2]\times 10^{44}$ erg s$^{-1}$ respectively, where numbers before [inside] the square brackets give the absorbed [de-absorbed] luminosities. Between 20 and 50 per cent of X-ray sources detected have a hard count ratio S/H$<$2.5, depending on the field (Table~\ref{tab:xraysample}). 

Optical selection has been somewhat less stringent, depending primarily on photometric weakness in archival catalogues, but affected by real-time observing conditions. Assuming an optical ($R$-band) to X-ray (0.5--7 keV) flux ratio $F_X/F_R=10$ or greater (which is true for the type 2 quasars in the field of A\,2390 (sources 15 and 18), the optical limit corresponding to our X-ray limits is $R\approx 25$. The other type 2 quasar A\,963\_15 has a ratio closer to $F_X/F_R=1$, due to the higher inferred X-ray obscuration, close to being Compton-thick. For sources with such a flux ratio, our X-ray limit implies an optical limit of $R\approx 22.5$. Approximately half of the hard X-ray sample have optical detections or limits fainter than the DSS.

Fifty-eight X-ray sources brighter than $F_{0.5-7}=2\times 10^{-15}$ \ergpspsqcm\ (average flux limit of the \c\ fields is $\sim 5\times 10^{-15}$ cgs) were imaged in the near-infrared and 56 were found to have counterparts brighter than $K\approx 20.7$, with K$_{\rm median}=17.7$. The two sources with no counterparts (A963\_2, a soft X-ray source and A963\_12, a very hard X-ray source) both have an X-ray:$K$ flux ratio $F_{\rm X}:F_{\rm K}\gtsim 10$ (arrows in Fig~\ref{fig:k_fx}). While redshift information is required to understand the nature of this difference, we note that other sources in our sample with an X-ray:$K$ flux ratio as large as 10 have a variety of X-ray hardness ratios: soft, hard and very hard. The strength of the X-ray flux comparative to that in the \k-band can thus be a signature of powerful accretion activity, either in type 1 or type 2 AGN.

We find that all very hard sources with $F_{\rm X}>3\times 10^{-14}$ cgs have \k$<$16.5 (hatched region of Fig~\ref{fig:sh_fx}). The luminosities implied from their redshifts (0.3--1.5) indicate that all (except one: A1835\_4) are powerful AGN with $L_{\rm X}>10^{44}$ erg s$^{-1}$. This is thus an ideal region to search for powerful, obscured accretion activity.

Photometric redshifts based on optical-NIR photometry resulted in 51 \zp-estimates. For four hard X-ray sources, good agreement was found between this estimate and the spectroscopic redshift (\zs) measurements. If there is a large amount of dust distributed similarly to the obscuring gas, the optical spectra and SEDs of obscured AGN can be dominated by their host galaxies, in which case template SEDs used for photometric-redshift fitting will be a good approximation of the underlying sources. The redshift distribution peaks at $z\approx 1$, with a slightly lower median redshift for the hard X-ray sources, as compared to the median redshift of objects with soft count ratios; however, results from a complete sample need to be compiled to verify this observation.

While the \k-magnitudes suggest the presence of massive black holes in the observed sample, the (albeit-weak) optical-NIR colour constraints are broadly consistent with the majority of hard sources having optical SEDs similar to those of galaxies, rather than unobscured quasars. 

Our selection procedure resulted in the clear detection of three type 2 quasars discussed in \S~\ref{sec:qso2}, and the identification of 9 other candidates. Two of the three clear identifications (A2390\_15, A2390\_18) have obscuring gas columns inferred from their X-ray spectra \nh$\approx$$2\times 10^{23}$ cm$^{-2}$, while the obscuration of the third source (A963\_15) is at least $10^{24}$ cm$^{-2}$. The inferred power which is being absorbed in all cases is $\sim 10^{45}$ erg s$^{-1}$ or greater.

\subsection{Other ways of selecting radio-quiet type 2 quasars}

There are currently no definitive selection criteria for locating the population of obscured AGN which are also weak in the radio. The paucity of X-ray selected AGN in recent observations at redshifts greater than $z=1$ over the expectation of background synthesis models complicates the issue since selection based on photometric dropouts will not work. Perhaps the most serious issue is the apparent lack of any discernible spectral line emission in the optical counterparts of many X-ray selected Seyferts. \citet{hasinger0301} finds that as many as 30 per cent of counterparts identified so far have galaxy-like spectra with weak or absent line emission. Whether or not this same fraction continues into the luminosity-regime of obscured quasars remains to be seen. At least one such source was identified by \citet{cowie02} and elaborated on by C02 (A2390\_15, discussed in \S~\ref{sec:qso2} of this paper). \citet{moran02} recently stressed the difficulties of dis-entangling the host galaxy and nuclear emission in typical optical spectra of Seyfert 2s obtained with slit-widths of $\sim$0.8 arcsec, which effectively weaken the equivalent-widths of lines. It is unclear, however, whether this argument can be extended to the regime of type 2 quasars, where the nuclei are more luminous by at least a factor of a few.

\citet{stern02} mention several techniques to search for type 2 quasars, including colour-selection and narrow-band imaging. Given the relatively-small redshifts of the newly-discovered X-ray population, however, a narrow-band Lyman alpha survey for sources at the characteristic redshift of $z=0.7$ would have to work at wavelengths close to 2070\AA, not an easy task. Narrow-band searches for strong forbidden line emission (e.g., [OIII] redshifted to the $I$-band), correlated with X-ray surveys, may prove useful for AGN that are not completely obscured.

Primary AGN radiation which is absorbed is likely to be reprocessed and emitted at longer wavelengths: current estimates of AGN contributions to this regime range from 20 per cent in the sub-mm \citep{almaini99} to 30 per cent or higher in the mid-infrared \citep{fadda01}, as determined from cosmic backgrounds and overlap with X-ray surveys. Thus, the next major space-based infrared mission, \sirtf, is likely to discover a large number of obscured AGN \citep[e.g., ][]{franceschini02, g03, lonsdale03}, even though it may be difficult to accurately distinguish emission from associated starbursts. However, until \sirtf\ begins full-time operation, a combination of X-ray and optical diagnostics are probably the best way to select powerful, obscured AGN.


\subsection{Number densities of type 2 quasars}
The local mass density in black holes can be used to place constraints on the space density of Seyferts and quasars, if one assumes a value for the radiative efficiency of accretion. While there is currently an uncertainty of at most a factor of a few in the fraction of the total mass accreted in obscured phases versus that accumulated through unobscured accretion, there seems to be little room for a large contribution by type 2 quasars \citep{f03}, and the bulk of this mass is likely accreted in AGN with Seyfert-like luminosities. 
 
In terms of areal density, if we assume (at face value) that our X-ray+optical selection procedure is effective and picks up a large fraction of all type 2 quasars, we find a total of 12 potential type 2 quasars in 12 \c\ fields. The flux limit for most fields is $\sim 5\times 10^{-15}$ erg s$^{-1}$ cm$^{-2}$, while the deepest limit is close to $\sim 2\times 10^{-15}$ erg s$^{-1}$ cm$^{-2}$. The derived density averages to $\sim$1 source per \c\ field covering $17 \times 17$ arcmin$^2$ (assuming the ACIS-I contiguous field-of-view, or four CCDs of ACIS-S) or $\sim$12 sources deg$^{-2}$, similar to the value derived by \citet{stern02}. Of course, we are probing a separate region of parameter space than those authors due to our larger sky coverage and correspondingly shallower flux limits (both CXO-52 described by them and CXOCDFS J033229.9-275106 described by \citealt{norman02} would have less than 10 net counts in our 30 ks X-ray exposures). This suggests that the true number density of type 2 quasars may have thus far been underestimated. Note that we have counted only one \c\ exposure per field and density variations of the order of 10 per cent are easily possible, if a different exposure is chosen for one of the fields observed twice (see the list of ACIS chips in Table~\ref{tab:xraysample}), or if, for instance, the shallowest and deepest fields are discarded in order to avoid biases.

This conclusion is partly supported by the deep field study of \citet{barger02}, who identify three narrow-line (in the optical spectra), powerful X-ray sources with $F_{0.5-8}>10^{-14}$ erg s$^{-1}$ cm$^{-2}$ and implied $L_{2-8}>10^{44}$ erg s$^{-1}$ in the Megasecond sample of \citet[][their sources 69, 184, 280]{brandt01}. Assuming their coverage to be 450 arcmin$^2$ (the largest area covered by them), this comes to about 2 such sources per \c\ ACIS-I field. 


The incompleteness of our X-ray follow-up, however, combined with large photometric-redshift uncertainties in several cases make our source density estimates uncertain. In the worst case, if we assume that the only \lq true\rq\ type 2 quasars are the three clear identifications discussed, their areal density would drop by a factor of four to 0.25 sources per ACIS-I field or 3 sources deg$^{-2}$. On the other hand, considering the fact that the majority of sources (more than 80 percent) were found on chips ACIS-S2 and ACIS S3 due to their higher sensitivities, one can derive an areal density that is higher by factor of two by including only 22 chips for 11 fields (the present Perseus analysis does not include S2 and S3). To summarize, despite uncertainties of a factor of a few, three sources deg$^{-2}$ can be considered as a strong lower limit to the density of type 2 quasars down to an X-ray flux $F_{0.5-7}\sim 5\times 10^{15}$ erg s$^{-1}$ cm$^{-2}$, and there is evidence from several data sets for a density higher than 10 deg$^{-2}$.

While complete, large-area samples will be needed to establish the true density, inferences from background synthesis models can be used refine the estimates, even though type 2 quasars do not dominate the X-ray background and thus provide a weak limit to their distribution. One recent model which has attempted to simultaneously fit all observational constraints including the XRB spectrum, log$N$-log$S$ and the redshift distribution of X-ray selected AGN is that of \citet[][see also \citealt{franceschini02}]{g03}. Above our deepest X-ray flux limit of $2\times 10^{-15}$ cgs, this model predicts 35 type 2 quasars and 115 type 1 quasars deg$^{-2}$ with $L_{2-10}>10^{44}$ erg s$^{-1}$, and similar numbers for the same luminosity limit in the 0.5--7 keV band. This translates to 2.9 type 2 quasars per $17\times 17$ arcmin$^2$ field. Above a flux of $5\times 10^{-15}$ cgs, 19 type 2 quasars deg$^{-2}$ are expected, not signicantly different from our optimistic estimates of 12 sources deg$^{-2}$. We note that the space density in the high luminosity regime is a rapidly decreasing function -- for X-ray luminosity $L_{2-10}>3\times 10^{44}$ erg s$^{-1}$, only three type 2 quasars are predicted by the model per square degree. 

Determination of the space densities of type 2 quasars will require constructing luminosity functions in various redshift intervals. Based on the (incomplete) redshift identifications in deep fields so far, \citet[][ see also \citealt{hasinger03, ueda03}]{cowie02} have determined a distinct evolution of AGN depending on {\em luminosity} as opposed to a distinction based on {\em obscuration} assumed by \citet[][]{g03}. Similar determinations from larger and complete samples at fainter fluxes are needed for an unbiased analysis of the distribution and evolution of type 2 quasars.\\ 

\noindent
Follow-up of sources found in very hard energy ranges could help select type 2 quasars effectively (e.g., \citealt{baldi02}). In the infrared, \sirtf\ surveys should provide another route for selecting these sources, based on photometry and broad-band SEDs covering the 10--100 \micron\ regime and extending coverage into the Compton-thick regime of obscuration. The difficulty of identifying emission lines in $z<1$ type 2 X-ray AGN may be overcome by searching for infrared emission lines. \citet[][]{spinoglio92} showed that \sirtf\ has the sensitivity to clearly detect bright forbidden line emission such as [OIII]~52\micron, [Ne~V]~24.2\micron\ and [S~IV]~10.5\micron\ from an AGN with line intensity $\sim 100$ times greater than that of NGC~1068, placed at $z\ltsim 0.3$ in a 1000~s exposure. Deep and very deep exposures could identify lines in weaker AGN and/or at higher redshift. Finally, the UKIRT Infrared Deep Sky Survey (UKIDSS), to be carried out with the Wide Field Infrared Camera (WFCAM; \citealt{wfcam}), will be a series of five public surveys of varying depth and area, the deepest of which should achieve $K_{\rm lim}=23$ over 1 deg$^2$. Cross-correlation of the NIR data with other wavebands can be expected to increase our understanding of the distribution of powerful, obscured AGN, including type 2 quasars.







\section{Acknowledgements}

PG would like to thank the Isaac Newton Trust and the Overseas Research Trust for support. CSC and ACF acknowledge financial support from the Royal
Society, and RMJ acknowledges support from PPARC. We are grateful to the anonymous referee for detailed and constructive comments.

This research owes a debt to the \c\ X-ray Observatory and related organizations and people for building and successfully operating the telescope. In addition, a large part of the data is based on observations carried out at the United Kingdom InfraRed Telescope. Archival databases maintained by the following telescopes and organizations have been made use of: the Digitized Sky Survey; the Isaac Newton Group of Telescopes; the Canada-France-Hawaii Telescope, the Anglo-Australian Telescope and the Hubble Space Telescope. Use was also made of the online NASA/IPAC Extragalactic Database. Steve Allen was the PI for the \c\ observations of Abell 963. Neil Trentham is thanked for making available archival optical data from the University of Hawaii 2.2m Telescope.

We also thank M. Bolzonella, J.-M. Miralles and R. Pell\'{o} for making HYPERZ publicly available.
\bibliographystyle{mnras}                       
\bibliography{mn-jour,gandhi_colour_astroph}

\onecolumn

\begin{table}
\begin{center}
\caption{Field List\label{tab:fieldlist}
}
\begin{tabular}{lccccccr}
\hline
Field$^a$                   &       RA$^b$  & DEC$^b$     & $z$     & Galactic N$_{\rm H}^c$ & Exposure$^d$ & \c\ Seq. Id. & Reference\\
                            &     J2000     & J2000       &         & $10^{20}$~cm$^{-2}$    & ks           &              & \\
\hline
Per [Perseus, Abell~426]    & 03:19:54.6    & $+$41:34:28 & 0.018   & 15.0                   &  5.0         & 800010       & \citet{permainref}\\
RXJ0821 [RX~J0820.9+0752]   & 08:21:02.4    & $+$07:51:47 & 0.110   & 2.12                   &  9.4         & 800115       & \citet{rxj0821mainref}\\
0902 [B2~0902+343]          & 09:05:30.1    & $+$34:07:57 & 3.395   & 2.31                   &  9.8         & 700212       & \citet{0902mainref}\\
IRAS09104 [IRAS~09104+4109] & 09:13:45.5    & $+$40:56:29 & 0.442   & 1.82                   &  9.1         & 800017       & \citet{iras09mainref}\\
A963 [Abell~963]            & 10:17:08.0    & $+$39:03:47 & 0.206   & 1.40                   & 36.3         & 800079       & e.g., \citet{allen02}\\
A1795 [Abell~1795]          & 13:48:49.2    & $+$26:36:29 & 0.063   & 1.19                   & 19.4         & 800001       & \citet{a1795mainref}\\
3C294 [3C~294]              & 14:06:44.1    & $+$34:11:24 & 1.786   & 1.21                   & 19.5         & 700204       & \citet{3c294mainref}\\
                            & 14:06:43.1    & $+$26:36:22 &         &                        & 70.0$^d$     & 800207       &\\
A1835 [Abell~1835]          & 14:01:02.0    & $+$02:52:40 & 0.252   & 2.30                   & 19.6         & 800003       & \citet{a1835mainref}\\
A2199 [Abell~2199]          & 16:28:32.2    & $+$39:33:35 & 0.030   & 0.88                   & 17.7         & 800005       & \citet{a2199mainref}\\
                            & 16:28:45.2    & $+$39:33:02 &         &                        & 15.1         & 800006       & \\
A2204 [Abell~2204]          & 16:32:44.4    & $+$05:33:25 & 0.152   & 5.67                   & 10.0         & 800007       & e.g., \citet{peres98}\\
MS2137 [MS~2137--2353]$^e$  & 21:40:14.0    & $-$23:40:43 & 0.313   & 3.55                   & 34.7         & 800104       & \citet{ms2137mainref}\\
A2390 [Abell~2390]          & 21:53:35.1    & $+$17:40:30 & 0.228   & 6.81                   &  9.8         & 800008       & \citet{a2390mainref}\\
                            & 21:53:37.6    & $+$17:41:17 &         &                        &  9.1         & 800009       & \\
\hline
\end{tabular}
\end{center}
~~\par
{\sl \footnotesize
$^a$Full designation given in square parentheses.\\
$^b$Coordinates are the pointing positions of \c.\\
$^c$Weighted average value from the maps of \citet{stark92}.\\
$^d$Exposure denotes the effective (good) exposure time. We note that 3C294 has been observed for an exposure time close to 200~ks. Here we present results of only the first 70~ks, as analysis of the full exposure was not available at the time of writing.\\
$^e$This field is drawn from the \c\ archive.\\
}
\end{table}

\begin{table}
\begin{center}
\caption{X-ray detections\label{tab:xraysample}
}
\begin{tabular}{llrr}
\hline
Field                      & Detections$^a$   & Hard (\%)$^b$& \k\ (\%)$^c$ \\
\hline
Per [800010]               & 18 [023]           & 9  (50)      & 1 (6)$^\dag$\\
RXJ0821                    & 21 [2367]           & 10 (48)      & 4 (20) \\
0902$+$343                 & 17 [67]           & 6  (35)      & 1 (6) \\
IRAS09104                  & 17 [2367]           & 5  (29)      & 3 (18) \\
A963                       & 61 [23567]           & 28 (46)      & 9 (15) \\
A1795                      & 24 [235678]           & 5  (21)      & 2 (9) \\
3C294 [700204]             & 31 [2367]           & 12 (39)      & 3 (10) \\
\hspace{1.07cm} [800207]  & 48 [67]           & 24 (50)      & 2 (5) \\
A1835                      & 27 [23678]           & 11 (41)      & 4 (15) \\
A2199 [800005]             & 33 [23567]           & 12 (36)      & 7 (21) \\
\hspace{1.07cm} [800006]  & 23 [2367]           & 7  (30)      & 4 (17) \\
A2204                      & 32 [235678]           & 13 (41)      & 2 (6) \\
MS2137                     & 35 [67]           & 9  (26)      & 4 (11) \\
A2390 [800008]             & 27 [23567]           & 8  (30)      & 11 (41) \\
\hspace{1.07cm} [800009]  & 23 [23678]           & 4  (17)      & 12 (52) \\
\hline
\end{tabular}
\end{center}
~~\par
{\sl \footnotesize
$^a$Column 2 (Detections) lists the total number of point sources detected in the \c\ observation. The numbers in square brackets denote the ACIS chips (given as ccd\_id numbers) which have been analysed for the present sample, e.g.: [267] implies chips 2, 6 and 7, i.e., ACIS-I2, S2 and S3.\\
$^b$Column 3 (Hard) lists the number of sources with S/H$<$2 (see \S~\ref{sec:selection}). The numbers in brackets represent the percentage of detections listed in column 2 with such a S/H limit.\\
$^c$Column 4 (\k) denotes the number of sources with \k-band detections presented herein, except for the one marked $^\dag$, which only has an optical detection and near infrared spectroscopy. Sources common to two separate observations of the same field have been included in both cases. The fraction of X-ray sources represented by this sample of \k-band detections is given in brackets as a percentage.
}
\end{table}

\begin{table}
\begin{center}
\caption{Archival Optical Observations Catalogue\label{tab:optical}
}
\begin{tabular}{lccccc}
\hline
Field                      & Filter$^a$ & Instrument & Date & Exposure  & Limiting Magnitude$^b$ \\
                           &        &            &      &    (s)    &   (3$\sigma$)      \\
\hline
Perseus                    & r      & INT WFC         & 2000 Oct 07 & 1000& 23.5 \\
                           & i      & INT WFC         & 2000 Oct 07 & 1000& 22.8 \\ 
RXJ0821                    & r      & INT WFC         & 2001 Jan 02 & 999 & 24.2 \\
                           & i      & INT WFC         & 2001 Jan 02 & 999 & 23.0 \\ 
0902$+$343                 & V      & INT WFC         & 1999 Jan 19 & 1079& 23.8 \\
                           & I      & INT WFC         & 1999 Jan 19 & 719 & 22.7 \\ 
A963                       & U      & WHT Prime Focus & 1994 Dec 09 & 1500& 24.0 \\ 
                           & B      & WHT Prime Focus & 1994 Dec 09 & 500 & 25.2 \\ 
                           & i      & INT WFC         & 2001 Jan 04 & 999 & 23.4 \\ 
A1795                      & i      & INT WFC         & 2000 May 01 & 1199& 22.8 \\ 
3C294                      & R      & HST WFPC2       & 1994 Apr 04 & 140 & 24.0 \\ 
                           & i      & INT WFC         & 2001 Mar 16 & 999 & 23.6 \\ 
A1835                      & R      & CFHT STIS2      & 1998 Feb 26 & 600 & 25.5 \\
                           & i      & INT WFC         & 2000 May 01 & 1199& 22.9 \\ 
A2199                      & B      & UH2.2m Tek 2048 & 1995 Nov 15 & 420 & 25.6 \\ 
                           & R      & UH2.2m Tek 2048 & 1995 Nov 15 & 300 & 24.6 \\ 
                           & i      & INT WFC         & 2000 May 01 & 1199& 22.9 \\ 
A2204                      & R      & INT Prime Focus & 1994 Jun 07 & 600 & 20.7 \\ 
MS2137-2353                & B      & AAT Prime Focus & 1993 Aug 12 & 600 & 25.4 \\
                           & I      & AAT Prime Focus & 1993 Aug 12 & 600 & 23.9\\
A2390                      & B      & WHT Prime Focus & 1994 Dec 08 & 500 & 23.6 \\ 
                           & R      & INT Prime Focus & 1994 Jun 06 & 600 & 23.4 \\ 
                           & I      & WHT Prime Focus & 1994 Dec 08 & 300 & 22.5 \\ 
                           & I      & CFHT Focam      & 1990 Oct 16 & 700 & 22.9 \\
\hline
\end{tabular}
\end{center}
~~\par
{\sl \footnotesize
$^a$Filters in small-caps typically refer to Gunn filters, while those in large-letters typically refer to Bessell filters.\\
$^b$Limiting Magnitude is calculated as a 3$\sigma$ deviation in a 3-arcsec diameter background aperture.\\
}
\end{table}

\begin{table}
\begin{center}
\caption{Near Infra Red Observations Log\label{tab:nirlog}
}
\begin{tabular}{lccr}
\hline
Field                      & Instrument   & Date             & Exposure\\
                           &              &                  & (s) \\
\hline
RXJ0821                    & UKIRT UFTI   & 29-30 Jan 2001   & 540\\
0902$+$343                 & UKIRT UFTI   & 30 Jan 2001      & 540\\
IRAS09104                  & UKIRT TUFTI  & 24-25 Feb 2000   & 540\\
A963                       & UKIRT UFTI   & 27-31 Jan 2001   & 540\\
A1795                      & UKIRT TUFTI  & 24-25 Feb 2000   & 540\\
3C294                      & UKIRT UFTI   & 27-28 Jan 2001   & 540\\
A1835                      & UKIRT UFTI   & 27-28 Jan 2001   & 540\\
                           & VLT ISAAC    & 28 Jun 2001      & 600\\
A2199                      & UKIRT UFTI   & 10-12 Aug 2000   & 540\\
A2204                      & UKIRT UFTI   & 11 Aug 2000      & 540\\
MS2137-2353                & VLT ISAAC    & 28 Jun 2001      & 720-900 (J); 600 (HK)\\
A2390                      & UKIRT UFTI   & 10-11 Aug 2000   & 540\\
\hline
\end{tabular}
\end{center}
~~\par
{\sl \footnotesize
The typical limiting magnitudes (in J, H and K respectively; 3$\sigma$ in a 3 arcsec diameter aperture) for the sample are as follows:\\
 UFTI: 21.4, 20.7, 20.0\\
 TUFTI: 21.6, 20.8, 20.1\\
 ISAAC: 22.6, 21.6, 20.7\\
The exposure in the last column refers to the typical time spent on a 9 point jitter pattern for one source. A few sources were multiply imaged in adjacent fields and the limiting magnitudes in these cases are correspondingly fainter.\\
The seeing in most cases ranged from $\approx$0.5--1.5 arcsec. Severely non-photometric data has been discarded from the current sample.
}
\end{table}

\begin{table}
\caption[Potential type 2 quasars]{Other potential type 2 quasars in our sample above an X-ray luminosity of $10^{44}$ erg s$^{-1}$ as inferred from Fig~\ref{fig:lx_z} or X-ray spectral information\label{tab:qso2}}
\begin{center}
\begin{tabular}{lccc}
\hline
Source   &        z        & $L_{0.5-7}$     &  S/H \\
         &                 & $10^{44}$ erg $^{-1}$ &                  \\
\hline
IRAS09\_3 & 1.43 [0.7, 2.1] &  2.7               & 2.5\p0.8\\
A963\_20 & 2.40 [2.0, 2.8] &  3.1                & 2.0\p0.7\\
A1795\_1 & 1.99 [0.8, 2.5] &  4.5                & 2.5\p0.9\\
A1795\_2 & 1.30 [0.8, 2.0] &  3.4                & 2.4\p0.4\\
3C294\_2 & 3.32 [1.4, 5.4] &  3.1                & 1.7\p1.1\\
3C294\_3 & 1.94 [0.9, 2.0] &  1.1                & 1.0\p0.2\\
A2199\_2 & 2.83 [1.0, 3.4] &  2.4                & 1.9\p1.2\\
MS2137\_4 & 3.33 [3.1, 3.4] &  1.7                & 1.6\p0.8\\
A2390\_12 & 2.28 [1.7, 3.3] &  2.9                & 2.0\p1.1\\
\\
\hline
Source   &        z        & $L_{0.5-7}$     &  \nh\\
         &                 & $10^{44}$ erg $^{-1}$ &       cm$^{-2}$\\
\hline
A963\_15  &      0.536?     & 8.0$^{\ddag}$ & $1.0\times 10^{24}$\\
A2390\_15 & 2.78 [2.1, 3.3] & 2.2$^{\ddag}$ & $2.2\times 10^{23}$\\
A2390\_18 &      1.467      & 10.0$^{\ddag}$ & $2.0\times 10^{23}$\\
\hline
\end{tabular}
\end{center}
~\par
{\sl \footnotesize
Column 1: Nine sources without X-ray spectral information are listed first in order of increasing right ascension, followed by the three sources with X-ray spectra.\\ 
Column 2: Numbers in brackets are 90 per cent \zphot\ confidence intervals. Numbers stated to three decimals places are spectroscopic measurements.\\
Column 3: The X-ray luminosities are the 0.5--7 keV luminosities under the assumptions of Fig~\ref{fig:lx_z}: $\Gamma=1.4$ and Galactic absorption only. The ones marked with a $^{\ddag}$ are measured intrinsic 2--10 keV luminosities from \c\ spectral fitting already discussed.\\
Column 4: For sources with X-ray spectral fitting, this column gives the measured $N_{\rm H}$ in cm$^{-2}$ (based on the fits with photon-index=2), and for the rest, the S/H ratios are stated.
}
\end{table}

\begin{landscape}


\begin{table}
\footnotesize
\caption{X-ray Source Catalogue \label{tab:xraycat}}
\begin{center}
\begin{tabular}{llccccrrcrcrcrccc}
Object & Seq. No.  [Chip] & RA    &$\Delta\alpha$& DEC   &$\Delta\delta$& Sig & Total & (err) & Soft  & (err) & Hard  & (err) & S/H & (err) & Flux             & log$L$\\
       &                  & J2000 & ($''$)       & J2000 & ($''$)       &     & (cts) &       & (cts) &       & (cts) &       &     &       & (cgs) & erg s$^{-1}$\\
 (1)   &    (2)           &  (3)  &     (4)      &  (5)  &    (6)       & (7) & (8)   & (9)   &  (10) & (11)  &  (12) &  (13) & (14)& (15)  & (16)             & (17) \\ 
\hline \hline 
Per\_2     & 800010 [I2] & 03:19:46.4 & (0.2) & 41:37:34 & (0.1) & 7.2 & 25.7 & (6.2) & 11.4 & (4.4) & 14.3 & (4.3) & 0.8 & (0.4) & 29.1 & 44.3 \\ 
\hline
RXJ0821\_1 & 800115 [S3] & 08:21:00.2 & (0.3) & 07:49:20 & (0.2) & 4.9 & 13.7 & (3.96) & 3.30  & (2.04) & 10.49 & (3.40) &0.32  & (0.22) & 10.0 & -- \\
RXJ0821\_2 & 800115 [S3] & 08:21:01.4 & (0.1) & 07:49:52 & (0.2) & 7.1 & 16.7 & (4.19) & 15.84 & (4.05) & 0.78  & (1.02) &20.26 & (27.06) & 12.6 & -- \\
RXJ0821\_3 & 800115 [S3] & 08:21:16.7 & (0.2) & 07:53:52 & (0.2) & 6.2 & 13.0 & (3.65) & 5.96  & (2.48) & 7.04  & (2.69) &0.85  & (0.48) & 9.8 & -- \\
RXJ0821\_4 & 800115 [S3] & 08:21:18.0 & (0.1) & 07:53:59 & (0.1) & 40.0 & 101.5 & (10.15) & 81.28 & (9.06) & 20.12 & (4.56) &4.04 & (1.02) & 76.5 & -- \\
\hline
0902\_1    & 700212 [S2] & 09:05:21.7 & (0.2) & 34:12:29 & (0.2) & 7.3 & 20.3 & (4.59) & 4.92 & (2.29) & 15.44 & (3.99) &0.32 & (0.17) & 14.7 & -- \\
\hline
IRAS09\_1  & 800017 [S2] & 09:13:41.0 & (0.4) & 41:03:15 & (0.3) & 22.4 & 52.2 & (7.3) & 38.8 & (6.2) &13.4 & (3.7) & 2.9 & (0.9) & 40.0 & -- \\
IRAS09\_2  & 800017 [S3] & 09:13:52.9 & (0.1) & 40:58:31 & (0.1) & 25.2   & 39.5 & (6.7) & 30.4 & (5.8) & 9.2 & (3.5) & 3.3 & (1.4) & 30.3 & --\\
IRAS09\_3  & 800017 [S2] & 09:13:57.5 & (0.2) & 40:59:39 & (0.1) & 25.0   & 47.5 & (6.9) & 33.8 & (5.8) & 13.8 & (3.7) & 2.5 & (0.8) & 36.4 & -- \\
\hline
A963\_1    & 800079 [S3] & 10:16:54.9 & (0.1) & 39:04:30  & (0.2) & 4.9 & 11.22 & (3.52) & 3.47 & (2.04) & 7.79 & (2.88) &0.45 & (0.31) & 2.2 & -- \\
A963\_2    & 800079 [S3] & 10:16:55.2 & (0.1) & 39:02:52  & (0.1) & 24.8 & 82.38 & (9.34) & 60.76 & (8.03) & 21.66 & (4.76) &2.81 & (0.72) & 15.9 & -- \\
A963\_4    & 800079 [S3] & 10:16:57.1 & (0.1) & 39:03:32  & (6.38) & 10.3 & 36.2 & (6.56) & 23.52 & (5.35) & 12.71 & (3.79) &1.85 & (0.69) & 7.0 & -- \\
A963\_6    & 800079 [S3] & 10:16:60.0 & (0.1) & 39:03:21  & (0.1) & 22.9 & 118.2 & (11.89) & 88.51 & (10.33) & 29.73 & (5.88) &2.98 & (0.68) & 22.8 & -- \\
A963\_7    & 800079 [S3] & 10:17:05.6 & (0.2) & 39:00:56  & (0.1) & 7.4 & 21.80 & (4.95) & 9.46 & (3.35) & 12.39 & (3.65) &0.76 & (0.35) & 4.2 & -- \\
A963\_8    & 800079 [S3] & 10:17:08.6 & (0.3) & 38:59:33  & (0.3) & 5.0 & 17.03 & (4.57) & 3.91 & (2.51) & 13.23 & (3.83) &0.30 & (0.21) & 3.3 & -- \\
A963\_10   & 800079 [S2] & 10:17:10.2 & (0.2) & 39:09:02  & (0.2) & 14.5 & 51.42 & (7.27) & 36.41 & (6.11) & 14.81 & (3.91) &2.46 & (0.77) & 9.9 & -- \\
A963\_12   & 800079 [S2] & 10:17:13.4 & (0.5) & 39:08:53  & (0.2) & 6.2 & 20.32 & (4.64) & 2.56 & (1.80) & 19.09 & (4.43) &0.13 & (0.10) & 3.9 & -- \\
A963\_13   & 800079 [S2] & 10:17:13.5 & (0.2) & 39:06:47  & (0.1) & 5.0 & 11.05 & (3.36) & 0.79 & (1.02) & 10.29 & (3.21) &0.08 & (0.10) & 2.1 & -- \\
A963\_15   & 800079 [S3] & 10:17:14.2 & (0.1) & 39:01:24  & (0.1) & 30.0 & 91.93 & (9.86) & 7.17 & (3.36) & 85.07 & (9.29) &0.08 & (0.04)& 54.5 & 43.7 \\
A963\_16   & 800079 [S2] & 10:17:17.5 & (0.4) & 39:10:15  & (0.3) & 5.6 & 17.09 & (4.37) & 5.95 & (2.57) & 11.27 & (3.55) &0.53 & (0.28) & 3.3 & -- \\
A963\_17   & 800079 [S2] & 10:17:21.0 & (0.4) & 39:09:38  & (0.4) & 5.2 & 11.99 & (3.65) & 1.64 & (1.48) & 10.52 & (3.36) &0.15 & (0.15) & 2.3 & -- \\
A963\_20   & 800079 [I2] & 10:18:02.4 & (0.8) & 39:00:07  & (0.7) & 8.2 & 55.48 & (8.62) & 36.55 & (6.62) & 18.49 & (5.50) &1.98 & (0.69) & 14.4 & -- \\
A963\_21   & 800079 [S3] & 10:16:51.2 & (0.1) & 39:04:19  & (0.1) & 19.8 & 49.62 & (7.15) & 39.70 & (6.37) & 9.93 & (3.24) &4.00 & (1.45) & 9.6 & -- \\
\hline
\end{tabular}
\end{center}
\par
{\sl \footnotesize
$^{(7)}$The significance given in column 7 is that obtained from WAVDETECT.\\
$^{(8)}$\lq Total\rq\ counts in column 8 refer to the 0.5--7 keV band.\\
$^{(9)}$\lq err\rq\ in columns 9, 11, 13 and 15 is the 1$\sigma$ error on the counts or the propogated error on the ratio.\\
$^{(14)}$S/H is the ratio of the counts in the Soft (0.5-2 keV; cols 10 and 11) band to those in the Hard (2-7 keV; cols 12 and 13) band.\\
$^{(16)}$Flux is stated for the Total X-ray band in units of $10^{-15}$ erg s$^{-1}$ cm$^{-2}$, calculated using PIMMS, assuming $\Gamma=1.4$ and only Galactic absorption at $z=0$.\\
$^{(17)}$Log of the rest-frame 0.5--7 keV luminosity calculated from the Total flux and spectral slope in the previous column, shown for sources with a spectroscopic redshift only.
}
\flushright {\sl continued...}
\end{table}


\addtocounter{table}{-1}
\begin{table}
\footnotesize
\caption{X-ray Source Catalogue (ctd)}
\begin{center}
\begin{tabular}{llccccrrcrcrcrccc}
Object & Seq. No.  [Chip] & RA    &$\Delta\alpha$& DEC   &$\Delta\delta$& Sig & Total & (err) & Soft  & (err) & Hard  & (err) & S/H & (err) & Flux             & log$L$\\
       &                  & J2000 & ($''$)       & J2000 & ($''$)       &     & (cts) &       & (cts) &       & (cts) &       &     &       & (cgs) & erg s$^{-1}$\\
 (1)   &    (2)           &  (3)  &     (4)      &  (5)  &    (6)       & (7) & (8)   & (9)   &  (10) & (11)  &  (12) &  (13) & (14)& (15)  & (16)             & (17) \\ 
\hline \hline
A1795\_1 & 800002 [S3] & 13:48:49.5 &0.1 &26:37:11 &0.1 & 16.7 & 84.9  & (13.8) & 61.0 & (12.1) & 23.9 & (6.7) &2.5 & (0.9) & 30.3 & -- \\
A1795\_2 & 800002 [S2] & 13:49:06.4 &0.1 &26:37:48 &0.1 & 21.6 & 158.8 & (13.2) & 112.4 & (11.1) & 46.4 & (7.1) &2.4 & (0.4) & 56.7 & -- \\
\hline
3C294\_1 & 800207 [S3] $^*$ & 14:06:29.2 & (0.1) & 34:12:36 & (0.1) & 26.53 & 90.14 & (9.93) & 10.87 & (3.69) & 79.15 & (9.21) &0.14 & (0.05) & 9.0 & -- \\
         & 700204 [S3] & 14:06:29.3 & (0.1) & 34:12:35 & (0.1) & 6.07  & 12.82 & (3.71) & 1.75  & (1.45) & 11.20 & (3.44) &0.16 & (0.14) & 4.6 & -- \\
3C294\_2 & 700204 [I3] & 14:06:36.3 & (0.7) & 34:01:16 & (1.0) & 4.53  & 15.69 & (4.59) & 9.89  & (3.44) & 5.68  & (3.01) &1.74 & (1.10) & 7.6 & --\\
3C294\_3 & 800207 [S3] $^*$ & 14:06:44.7 & (0.1) & 34:11:37 & (0.1) & 20.44 & 78.72 & (9.21) & 38.32 & (6.41) & 40.39 & (6.61) &0.95 & (0.22) & 7.9 & -- \\
         & 700204 [S3] & 14:06:44.8 & (0.1) & 34:11:36 & (0.1) & 19.71 & 41.70 & (6.50) & 25.81 & (5.11) & 15.88 & (4.01) &1.62 & (0.52) & 15.0& -- \\
\hline
A1835\_1$^a$ & 800003 [S3] & 14:01:07.0 & (0.3) &  02:49:34 & (0.3) & 5.82 & 18.11 & (4.56) & 5.65 & (2.70) & 13.70 & (3.82) & 0.41 & (0.23) & 6.3 & 43.5\\
A1835\_2 & 800003 [I2] & 14:01:49.7 & (1.1) &  02:48:38 & (0.8) & 4.06 & 18.81 & (7.49) & 14.14 & (6.12) & 4.54 & (2.12) & 3.11 & (1.98) & 9.0 & 44.7\\
A1835\_3     & 800003 [I3] & 14:01:30.5  & (0.8) & 02:45:29  & (0.4) & 5.89 & 23.04 & (5.33) & 10.77 & (3.59) & 12.38 & (3.96) &0.87 & (0.40) & 11.2 & --\\
A1835\_4     & 800003 [I2] & 14:01:44.9  & (0.6) & 02:53:33  & (0.5) & 10.76 & 65.89 & (8.90) & 40.60 & (6.86) & 67.76 & (9.02) &0.60 & (0.13) & 32.0 & 42.7\\
\hline
A2199\_1 & 800005 [S2] & 16:28:07.2 & (0.6) & 39:36:03 & (0.5)   & 6.62 & 36.15 & (7.47) & 22.13 & (6.01) & 14.32 & (4.46) &1.54 & (0.64) & 14.3 & -- \\
A2199\_2 & 800005 [S2] & 16:28:10.1 & (0.7) & 39:35:13 & (0.3)   & 4.81 & 19.96 & (5.88) & 13.12 & (4.88) & 6.78 & (3.30) &1.94 & (1.19) & 7.9 & -- \\
A2199\_3 & 800005 [S2] & 16:28:11.5 & (0.4) & 39:37:03 & (0.2)   & 6.36 & 19.79 & (4.98) & 14.59 & (4.28) & 5.21 & (2.56) &2.80 & (1.60) & 7.8 & -- \\
A2199\_4 & 800005 [S2] & 16:28:11.6 & (0.6) & 39:37:12 & (0.4)   & 6.50 & 30.26 & (7.08) & 19.51 & (5.80) & 10.82 & (4.06) &1.80 & (0.86) & 11.9 & --\\
A2199\_5 & 800005 [S2] & 16:28:12.1 & (0.4) & 39:36:10 & (0.4)   & 7.81 & 49.28 & (8.35) & 26.91 & (6.47) & 21.44 & (5.19) &1.25 & (0.43) & 19.4 & -- \\
A2199\_6 & 800006 [S3] & 16:28:23.3 & (0.2) & 39:34:12 & (0.2)   & 3.32 & 10.87 & (4.52) & 10.40 & (4.26) & 0.47 & (1.48) & 22.31 & (71.56) & 4.3 & 39.9 \\
A2199\_7 & 800005 [S3]      & 16:28:24.0 & (0.1) & 39:33:23 & (0.1)   & 12.13 & 48.37  & (7.90)  & 28.17  & (6.29)  & 20.23 & (4.77) &1.39 & (0.45) & 19.1 & --\\
         & 800006 [S3] $^*$ & 16:28:24.1 & (0.2) & 39:33:22 & (0.2)   & 8.43  & 39.89  & (7.83)  & 33.05  & (7.04)  &  6.86 & (3.43) &4.82 & (2.62) & 18.5 & --\\
A2199\_8 & 800005 [S3]      & 16:28:26.1 & (0.0) & 39:33:54 & (0.0)   & 56.26 & 276.92 & (17.16) & 221.88 & (15.36) & 55.07 & (7.66) &4.03 & (0.63) & 109.0 & -- \\
         & 800006 [S3] $^*$ & 16:28:26.1 & (0.1) & 39:33:55 & (0.1)   & 27.06 & 159.37 & (13.71) & 129.22 & (12.34) & 30.22 & (5.98) &4.28 & (0.94) & 73.5 & -- \\

A2199\_9$^b$ & 800006 [I3] & 16:28:27.6 & (0.6) & 39:23:43 & (0.6)   & 8.11  & 49.57  & (8.45)  & 35.20  & (7.02)  & 14.39 & (4.71) &2.45 & (0.94) & 30.8 & -- \\
\hline
A2204\_1 & 800007 [S2] & 16:32:37.2 & (0.2) &  05:29:42 & (0.2) & 14.40 & 38.08 & (6.30) & 5.17 & (2.51) & 33.21 & (5.81) &0.16 & (0.08) & 36.2 & -- \\
A2204\_2 & 800007 [I3] & 16:31:58.6 & (0.9) &  05:39:03 & (1.0) & 6.00 & 46.93 & (7.16) & 35.17 & (6.13) & 11.49 & (3.67) & 3.06 & (1.11) & 44.7 & -- \\
\hline
MS2137\_1 & 800104 [S2] & 21:40:29.3 & (0.7) & --23:47:41 & (0.9) & 3.32 & 11.98 & (4.56) & 6.92 & (3.19) & 6.50 & (3.47) &1.07 & (0.75) & 3.3 & 43.8\\
MS2137\_2 & 800104 [S2] & 21:40:30.7 & (1.0) & --23:47:46 & (0.7) & 4.15 & 13.30 & (4.43) & 8.17 & (3.19) & 5.21 & (3.11) &1.57 & (1.12) & 3.6 & --\\
MS2137\_3 & 800104 [S3] & 21:40:12.2 & (0.5) & --23:35:03 & (0.5) & 3.92 & 12.73 & (4.30) & 3.72 & (2.57) & 9.11 & (3.47) &0.41 & (0.32) & 2.5 & --\\
MS2137\_4 & 800104 [S3] & 21:39:52.4 & (0.4) & --23:37:37 & (0.3) & 5.72 & 20.59 & (5.18) & 12.89 & (4.09) & 8.08 & (3.26) &1.59 & (0.82) & 4.1 & -- \\
\hline
\end{tabular}
\end{center}
~\par
{\sl
$^a$This is source B3 in \citet{c01} and its X-ray coordinates were erroneously placed 9 arcsec away therein.\\
$^b$This is source C1 in \citet{c01} and the initial \c\ astrometric offsets were not removed for this source therein.\\
$^*$Several sources have detections on two \c\ exposures and for these, a $^*$ denotes the detection that is plotted in Figs. 3, 4, 5 and 7.\\
}
\flushright
{\sl \footnotesize continued...}
\end{table}

\addtocounter{table}{-1}
\begin{table}
\footnotesize
\caption{X-ray Source Catalogue (ctd)}
\begin{center}
\begin{tabular}{llccccrrcrcrcrccc}
Object & Seq. No.  [Chip] & RA    &$\Delta\alpha$& DEC   &$\Delta\delta$& Sig & Total & (err) & Soft  & (err) & Hard  & (err) & S/H & (err) & Flux             & log$L$\\
       &                  & J2000 & ($''$)       & J2000 & ($''$)       &     & (cts) &       & (cts) &       & (cts) &       &     &       & (cgs) & erg s$^{-1}$\\
 (1)   &    (2)           &  (3)  &     (4)      &  (5)  &    (6)       & (7) & (8)   & (9)   &  (10) & (11)  &  (12) &  (13) & (14)& (15)  & (16)             & (17) \\ 
\hline \hline             
A2390\_4 & 800008 [I3] & --  & --  &-- &--  & --  & $<$4.63 &--&$<$4.58 & --  & $<$0.41 & -- &-- & -- & -- & -- \\
         & 800009 [I3] $^*$ & 21:53:00.9 & (0.4) & 17:44:52 & (0.7) & 4.20 & 11.46 & (3.58) & 7.88 & (2.90) & $<$4.18 & -- &$>$1.89  & -- & 9.4 & 43.7\\
A2390\_6 & 800008 [S3] $^*$ & 21:53:20.7 & (0.1) & 17:41:19 & (0.1) & 19.22 & 41.34 & (6.48) & 29.19 & (5.44) & 12.18 & (3.51) &2.40 & (0.82) & 36.1 & 44.1\\
         & 800009 [S3] & 21:53:20.7 & (0.1) & 17:41:19 & (0.1) & 23.77 & 59.77 & (7.79) & 50.51 & (7.15) & 9.23 & (3.09) & 5.47 & (1.99) & 48.4 & 44.3\\
A2390\_8 & 800008 [S3] $^\dag$ & 21:53:23.8 & (0.1) & 17:39:30 & (0.1) & 9.04 & 19.74 & (4.52) & 18.92 & (4.40) & $<$2.04 & --  &$>$9.27  & -- & 13.9 & 43.4\\
         & 800009 [S3] $^*$   & 21:53:23.8 & (0.1) & 17:39:30 & (0.2) & 7.97 & 23.25 & (4.87) & 19.21 & (4.42) & $<$4.08 & --  &$>$4.71  & -- & 15.2 & 43.5 \\
A2390\_10 & 800008 [S3] & -- & -- &-- & -- & --  & 6.53  & (3.19) &$<$5.62 & --  & $<$3.02 & --  & -- & -- & -- & --\\
          & 800009 [S3] $^*$ & 21:53:25.3 & (0.3) & 17:43:22 & (0.2) & 5.30 & 13.04 & (3.79) & 13.44 & (3.79) & $<$0.43 & --  &$>$31.26 & -- & 10.7 & 42.1\\
A2390\_12 & 800009 [S4] & 21:53:29.0 & (0.5) & 17:48:54 & (0.6) & 5.02 & 18.21 & (4.65) & 12.28 & (3.79) & 6.04 & (2.69) &2.03 & (1.10) & 14.9 & -- \\
A2390\_13 & 800009 [S3] & 21:53:31.60 & (0.3) & 17:47:04 & (0.3) & 8.41 & 21.61 & (4.75) & 18.78 & (4.39) & $<$3.62 & --  &$>$5.9 & -- & 17.7 & 44.2\\
A2390\_14 & 800008 [S2] & -- &--  &-- &--  & --  & $<$3.12 &--  &$<$0.40 & -- & $<$3.14 &--  &-- & -- & -- & --\\
          & 800009 [S2] $^*$ & 21:53:32.8 & (0.5) & 17:32:23 & (0.8) & 4.90 & 15.81 & (4.34) & 10.97 & (3.54) & $<$4.98 & -- &$>$2.20 & -- & 13.0 & --\\
A2390\_15$^c$ & 800008 [S3]  & 21:53:33.2 & (0.1) & 17:42:10 & (0.1) & 11.30 & 36.60 & (6.83) & 20.15 & (5.24) & 16.47 & (4.38) &1.22 & (0.46) & 28.1 & --\\
              & 800009 [S3] $^*$ & 21:53:33.2 & (0.1) & 17:42:10 & (0.1) & 11.38 & 40.51 & (7.03) & 23.19 & (5.42) & 17.33 & (4.49) &1.30 & (0.47) & 31.1 & --\\
A2390\_16 & 800008 [S2] $^*$ & 21:53:33.6 & (0.1) & 17:37:35 & (0.1) & 15.44 & 31.59 & (5.72) & 23.79 & (4.95) & 7.77 & (2.87) &3.06 & (1.30) & 22.9 & 44.4\\
          & 800009 [S2] & 21:53:33.6 & (0.1) & 17:37:35 & (0.2) & 13.84 & 27.33 & (5.28) & 21.34 & (4.64) & 5.93 & (2.50) &3.60 & (1.71) & 19.8 & 44.3\\
A2390\_17$^d$ & 800008 [S3] & 21:53:33.8 & (0.1) & 17:41:15 & (0.1) & 5.35 & 15.61 & (4.26) & 15.15 & (4.13) & $<$2.02& --  &$>$7.49 & -- & 9.9  & 43.4\\
              & 800009 [S3] $^*$  & 21:53:33.7 & (0.1) & 17:41:15 & (0.1) & 6.24 & 23.77 & (5.31) & 19.89 & (4.81) & $<$4.50& --  &$>$4.42 & -- & 15.0 & 43.6\\
A2390\_18$^e$ & 800008 [S3] $^*$ & 21:53:34.0 & (0.1) & 17:42:41 & (0.1) & 20.50 & 71.96 & (9.05) & 24.52 & (5.61) & 47.52 & (7.11) &0.52 & (0.14) & 104.9 & 44.6\\
              & 800009 [S3] & 21:53:34.0 & (0.1) & 17:42:41 & (0.1) & 22.84 & 76.51 & (9.14) & 29.69 & (5.86) & 46.88 & (7.02) &0.63 & (0.16) & 111.5 & 44.6\\
A2390\_19 & 800008 [S2] & 21:53:34.8 & (0.1) & 17:36:31 & (0.1) & 49.44 & 115.39 & (10.77) & 79.65 & (8.95) & 35.61 & (5.99) &2.24 & (0.45) & 86.1 & 44.9 \\
          & 800009 [S2] $^*$ & 21:53:34.8 & (0.1) & 17:36:31 & (0.1) & 49.88 & 127.53 & (11.35) & 95.34 & (9.81) & 31.80 & (5.67) &3.00 & (0.62) & 95.2 & 45.0 \\
A2390\_20 & 800008 [S3] $^*$ & 21:53:40.8 & (0.1) & 17:44:15 & (0.1) & 54.56 & 153.88 & (12.48) & 61.98 & (7.95) & 92.49 & (9.65) &0.67 & (0.11) & 114.0 & 43.5\\
          & 800009 [S3] & 21:53:40.8 & (0.1) & 17:44:16 & (0.1) & 54.39 & 139.67 & (11.87) & 61.97 & (7.91) & 78.06 & (8.86) &0.79 & (0.14) & 103.5 & 43.4\\
A2390\_24 & 800008 [S3] $^*$ & 21:53:45.6 & (0.2) & 17:41:48 & (0.1) & 9.75 & 31.59 & (6.37) & 13.61 & (4.45) & 19.14 & (4.67) &0.71 & (0.29) & 17.7 & 42.3\\
          & 800009 [S3]  & 21:53:45.5 & (0.1) & 17:41:47 & (0.1) & 6.61 & 15.65 & (4.18) &  8.18 & (3.04) &  7.49 & (2.87) &1.09 & (0.58) &  8.8 & 42.0\\
A2390\_26 & 800008 [S2] $^{\dag *}$ & 21:53:48.0 & (0.3) & 17:37:56 & (0.4) &  6.58 & 13.42 & (3.84) & 11.83 & (3.54) & $<$2.08 &  --    &$>$5.69 & --     & 10.8 & 44.0\\
          & 800009 [S2]        & 21:53:48.0 & (0.1) & 17:37:56 & (0.2) & 16.40 & 36.12 & (6.07) & 29.85 & (5.51) &    6.14 & (2.53) &4.86    & (2.19) & 29.2 & 44.4\\
A2390\_27 & 800008 [S3]  & 21:53:48.0 & (0.2) & 17:42:48 &  (0.2) & 10.80 & 23.08 & (4.91) & 15.94 & (4.09) & 7.17 & (2.71) &2.23 & (1.02) & 13.5 & --\\
          & 800009 [S3] $^*$ & 21:53:48.0 & (0.1) & 17:42:47 &  (0.1) & 17.05 & 52.96 & (7.34) & 45.83 & (6.82) & 7.10 & (2.70) &6.45 & (2.63) & 31.5 & --\\
A2390\_28 & 800008 [S3] & 21:53:48.8 & (0.1) & 17:42:00 & (0.2) & 13.78 & 30.67 & (5.70) & 24.96 & (5.11) & 5.70    & (2.52) &4.38     & (2.13) & 42.5 & 44.2\\
          & 800009 [S3] $^*$ & 21:53:48.9 & (0.2) & 17:42:00 & (0.2) &  5.86 & 14.06 & (4.08) & 10.79 & (3.53) & $<$4.10 & --     & $>$2.63 & --     & 18.1 & 43.8\\
\hline\hline
\end{tabular}
\end{center}
~\par
{\sl
$^c$This source is CXOU J215333.2+174209 (source 2) in \citet{f00}, source 3 (updated to CXOU J215333.2+174211) in \citet{cowie01} and source B in \citet{wfg}.\\
$^d$This source is CXOU J215333.8+174113 (source 3) in \citeauthor{f00} and source 2 (updated to CXOU J215333.8+174116) in \citeauthor{cowie01}.\\
$^e$This source is CXOU J215334.0+174240 (source 1) in \citeauthor{f00}, source A in \citet{wfg} and source 1 (updated to CXOU J215334.0+174242) in \citeauthor{cowie01}.\\
$^\dag$These sources are close to the edge of the chip,
where dither may affect the counts and S/H ratios.\\
$^*$As mentioned before, several sources have detections on two \c\ exposures and for these, a $^*$ denotes the detection that is plotted in Figs. 3, 4, 5 and 7.\\
}
\end{table}





\begin{table}
\footnotesize
\caption{Follow-up Photometry\label{tab:mags}}
\begin{center}
\begin{tabular}{lcccccccccc}
Object  & RA & Dec &U &B &R &i & J & H & K\\
        &J2000&J2000& &  &  &  &   &   &  \\
\hline\hline
Per\_2$^{P}$& 03:19:46.4        & +41:37:34         & -- &  --          & 21.05\p0.02$^r$& 20.07\p0.04 &     --      &$<$17.5$^{P}$ &  $<$16.4$^{P}$ \\
\hline
RXJ0821\_1 & 08:21:00.2$^{opt}$ & +07:49:19$^{opt}$ & -- &  --          & 24.23\p0.18$^r$& 22.77\p0.13 & 20.26\p0.14 & 19.57\p0.13 & 18.87\p0.09\\
RXJ0821\_2 & 08:21:01.5$^{opt}$ & +07:49:51$^{opt}$ & -- &  --          & 20.84\p0.01$^r$& 20.52\p0.02 & 19.92\p0.08 & 19.87\p0.11 & 18.74\p0.06\\
RXJ0821\_3 & 08:21:16.7         & +07:53:51         & -- &  --          & 21.52\p0.04$^r$& 20.80\p0.04 & $>$20.9    & 19.45\p0.09 & 18.98\p0.09\\
RXJ0821\_4 & 08:21:17.9         & +07:53:58         & -- &  --          & 20.89\p0.01$^r$& 19.90\p0.02 & 19.13\p0.07& 18.27\p0.04 & 17.49\p0.03\\
\hline
0902\_1   & 09:05:21.7$^{opt}$ & +34:12:30$^{opt}$ & -- & $>$22.5$^D$  & $>$23.8$^V$      & 22.47\p0.11$^{I}$ & 20.64\p0.13 & 19.73\p0.12 & 18.75\p0.09\\
\hline
IRAS09\_1$^\dag$ & 09:13:40.9 & +41:03:14 & -- & $>$22.5 &$>$21     & -- & 19.63$\pm$0.07 &18.86$\pm$0.05 &17.81$\pm$0.05\\
IRAS09\_2$^\dag$ & 09:13:52.9 & +40:58:30 & -- & $>$22.5 &$>$21     & -- & 19.76$\pm$0.07 &18.60$\pm$0.05 &17.47$\pm$0.05\\
IRAS09\_3$^\dag$ & 09:13:57.6 & +40:59:39 & -- & 21.5\p0.5 &20.0\p0.5 & -- & 18.63$\pm$0.05 &17.88$\pm$0.05 &17.04$\pm$0.05\\
\hline
A963\_1  & 10:16:54.7 & +39:04:31  & $>$24.0 & $>$25.2 & $>$21.0$^D$ & $>$23.4 & $>$21.4 & $>$20.7 & 19.82\p0.3\\
A963\_2  &  -- & --  & $>$24.0 & $>$25.2 & $>$21.0$^D$ & $>$23.4 & $>$20.9 & $>$20.7 & $>$20.0\\
A963\_4  & 10:16:57.0 & +39:03:32 & $>$24.0 & $>$25.2 & $>$21.0$^D$ & 22.10\p0.13 & 20.31\p0.11 & 19.22\p0.06 & 18.27\p0.04\\
A963\_6  & 10:16:59.9 & +39:03:21 & 23.92\p0.35 & 24.03\p0.11 & $>$21.0$^D$ & -- & 19.83\p0.15 & 19.75\p0.09 & 18.34\p0.05\\
A963\_7  & 10:17:05.5 & +39:00:56 & 23.14\p0.17  & 22.59\p0.04 & 20.68\p0.5$^D$ & 20.01\p0.03 & 19.14\p0.04 & 18.14\p0.03 & 17.39\p0.03 \\
A963\_8  & 10:17:08.5 & +38:59:33 & $>$24.0 & 24.44\p0.14 & $>$21.0$^D$ & 22.16\p0.23 & 20.26\p0.15 & 19.19\p0.10 & 18.15\p0.10 \\
A963\_10$^{3''}$ & 10:17:10.0 & +39:09:02 & -- & 21.0\p0.5$^D$ & 18.5\p0.5$^D$& 18.19\p0.03& 17.19\p0.02 & 16.48\p0.02 & 15.81\p0.04 \\
A963\_12 & --  & -- & $>$24.0 & $>$25.2 & $>$21.0$^D$ & $>$23.4 & $>$21.9 & $>$21.3 & $>$20.9\\
A963\_13 & 10:17:13.3 & +39:06:47 & -- & $>$22.5$^D$ & $>$21.0$^D$  & 20.90\p0.06 & 20.08\p0.20 & 19.02\p0.10 & 17.93\p0.08\\
A963\_15 & 10:17:14.1 & +39:01:25 & 23.13\p0.4 & 22.83\p0.04 & 19.75\p0.50 & 19.47\p0.03$^D$ & 18.3\p0.04 & 17.30\p0.02 & 16.47\p0.02\\
A963\_16 & 10:17:17.7 & +39:10:15 & -- & $>$22.5$^D$ & $>$21.0$^D$  & 21.96\p0.15 & $>$21.9 & $>$21.3 & 19.95\p0.3\\
A963\_17$^{3''}$ & 10:17:20.9 & +39:09:39 & -- & $>$22.5$^D$ & 20.97\p0.50 & 20.14\p0.03 & 19.06\p0.04 & 18.23\p0.04 & 17.10\p0.05\\
A963\_20 & 10:18:02.5 & +39:00:09 & -- & 22.5\p0.5$^D$ & $>$21.0$^D$ & 21.42\p0.03 & -- & 19.47\p0.11 & 19.14\p0.11\\
A963\_21 & 10:16:51.1 & +39:04:19 & $>$24.0 & $>$25.2  & $>$21.0$^D$ & $>$23.4 & $>$21.4 & $>$20.7 & 19.90\p0.3\\
\hline
\end{tabular}
\end{center}
~\par
{\sl \footnotesize
$^{P}$This source is designated P2 in \citet{c01}, the coordinates denote the TUFTI pointing position, and the magnitudes represent lower limits to the flux (upper limits to the mags) measured from spectroscopy.\\
$^{opt}$The coordinates for all sources were measured on the NIR images, except for these, which were measured on the optical images.\\
$^D$Magnitude estimated from the DSS.\\
$^V$This is a Harris $V$-band magnitude.\\
$^r$Sloan system $r$-band magnitude.\\
$^I$RGO $I$-band filter magnitude.\\
$^\dag$The NIR magnitudes for these sources were obtained with IRCAM3/TUFTI.\\
$^{3''}$The magnitudes for these sources was measured in a 3 arcsec aperture in all filters and corrected for seeing appropriately.
}
\flushright
{\sl \footnotesize continued...}
\end{table}

\addtocounter{table}{-1}
\begin{table}
\small
\caption{Follow-up Photometry (ctd)}
\begin{center}
\begin{tabular}{lccccccccc}
Object  & RA & Dec &B &R &i & J & H & K\\
        &J2000&J2000& &  &  &  &   &   &  \\
\hline\hline
A1795\_1$^\dag$ & 13:48:49.5 & 26:37:12 & $>$22.5 & $>$21 & -- & 22.43$\pm$0.23 &21.13$\pm$0.13 &19.99$\pm$0.13\\
A1795\_2$^\dag$ & 13:48:06.4 & 26:37:48 & $>$22.5 & $>$21 & -- & 18.76$\pm$0.05 &17.98$\pm$0.05 &17.01$\pm$0.05\\
\hline
3C294\_1 & 14:06:29.3$^{opt}$ & +34:12:36$^{opt}$ & 22.5\p0.5$^D$& 19.5\p0.3$^D$  & 19.19\p0.02     & 17.73\p0.03 & 16.94\p0.02 & 16.18\p0.02\\
3C294\_2 & 14:06:36.5 & +34:01:16                 & $>$22.5$^D$  & $>$21$^D$      & 22.13\p0.13$^F$ & $>$21.3     & $>$20.6     & 20.89\p0.26\\
3C294\_3 & 14:06:44.8 & +34:11:37                 & $>$22.5$^D$  & 23.97\p0.12$^H$& $>$23.6             & --          & --     & 19.57\p0.12\\
\hline
A1835\_1  & 14:01:06.9          & 02:49:33 & $>$22.5$^D$ & 24.04\p0.14$^C$ & $>$22.9$^z$     & 21.14\p0.13 & 20.81\p0.11 & 19.58\p0.11\\
A1835\_2  & 14:01:49.9          & 02:48:36 & $>$22.5$^D$ & 21.0\p0.5$^D$   & 20.87\p0.10$^z$ & 19.90\p0.08 & 19.16\p0.06 & 18.16\p0.04\\
A1835\_3  & 14:01:30.6$^{opt}$  & 02:45:31$^{opt}$ & 22.5\p0.5$^D$& $>$21$^D$          & 20.14\p0.04 & 18.71\p0.05 & 17.70\p0.04 & 16.64\p0.02\\
A1835\_4  & 14:01:45.0          & 02:53:33         & 19.6\p0.5$^D$& 17.2\p0.5$^D$      & 17.29\p0.01 & 16.17\p0.01 & 15.24\p0.01 & 14.36\p0.01\\
\hline
A2199\_1 & 16:28:07.2 & 39:36:02 & 23.06\p0.02$^{UH}$ & 21.64\p0.02$^{UH}$ & 20.38\p0.04 & 18.76\p0.03 & 18.27\p0.03 & 17.00\p0.02\\
A2199\_2 & 16:28:10.2 & 39:35:12 & 24.02\p0.02$^{UH}$ & 22.79\p0.02$^{UH}$ & 22.22\p0.10 & 21.09\p0.08 & 21.15\p0.16 & 20.27\p0.12\\
A2199\_3 & 16:28:11.5 & 39:37:02 & 25.28\p0.02$^{UH}$ & 24.20\p0.02$^{UH}$ & 23.26\p0.15 & $>$20.3     & 19.78\p0.21 & 19.07\p0.18\\
A2199\_4 & 16:28:11.6 & 39:37:11 & 25.00\p0.02$^{UH}$ & 23.74\p0.02$^{UH}$ & 24.31\p0.38 & $>$20.3     & $>$20.1     & 19.72\p0.28\\
A2199\_5 & 16:28:12.3 & 39:36:12 & 24.17\p0.02$^{UH}$ & 22.69\p0.02$^{UH}$ & 21.04\p0.07 & 19.35\p0.04 & 18.40\p0.03 & 17.40\p0.02\\
A2199\_6 & 16:28:23.3 & 39:34:13 & 16.40\p0.02$^{UH}$ & 14.86\p0.02$^{UH}$ & 13.62\p0.01 & 13.16\p0.01 & 12.46\p0.01 & 12.13\p0.01\\
A2199\_7 & 16:28:24.0 & 39:33:22 & 23.83\p0.02$^{UH}$ & 22.48\p0.02$^{UH}$ & 21.78\p0.06 & 20.53\p0.10 & 19.45\p0.09 & 17.99\p0.04\\
A2199\_8 & 16:28:26.1 & 39:33:53 & 20.30\p0.01$^{UH}$ & 19.41\p0.02$^{UH}$ & 19.04\p0.01 & 18.63\p0.04 & 17.76\p0.03 & 16.94\p0.02\\
A2199\_9 & 16:28:27.8 & 39:23:43 & --                 & --                 & 20.65\p0.03 & 19.20\p0.06 & 18.01\p0.04 & 17.42\p0.03\\
\hline
A2204\_1  & 16:32:37.2 & $+$05:29:42 & $>$22.5$^D$ & 19.60\p0.23 & -- & 18.35\p0.07 & 17.25\p0.04 & 16.38\p0.02\\
A2204\_2  & 16:31:58.6 & $+$05:39:03 & $>$22.5$^D$ & 20.0\p0.4$^D$      & -- & 19.06\p0.08 & 18.13\p0.05 & 17.36\p0.03\\
\hline
MS2137\_1$^\ddag$ & 21:40:29.3 & --23:47:41 & $>$22.5$^D$ & $>$21.0$^D$        & --              & 22.00\p0.12 & 20.73\p0.13 & 19.44\p0.09\\
MS2137\_2$^\ddag$ & 21:40:30.7 & --23:47:46 & $>$22.5$^D$ & $>$21.0$^D$        & --              & 22.28\p0.17 & 21.87\p0.16 & 20.46\p0.12\\
MS2137\_3$^\ddag$ & 21:40:12.2 & --23:35:03 & 24.54\p0.20 & $>$21.0$^D$    & 20.36\p0.03 & 19.45\p0.02 & 18.40\p0.04 & 17.76\p0.02\\%
MS2137\_4$^\ddag$ & 21:39:52.4 & --23:37:37 & $>$25.4 & $>$21.0$^D$        & 22.24\p0.12 & 21.55\p0.12 & 20.02\p0.07 & 19.02\p0.05\\
\hline
\end{tabular}
\end{center}
~\par
{\sl \footnotesize
$^F$These fields could not be de-fringed and background estimation was thus performed manually (see text).\\
$^H$This is an HST F702 magnitude.\\
$^C$This magnitude is based on a single flux standard observation. Cannot resolve the two components of this source in $R$.\\
$^z$Photometric standard star unavailable. Standard zero-points assumed, since observing conditions were reported to be photometric by the Carlsberg Meridian Telescope Extinction Monitor (http://www.ast.cam.ac.uk/$^{\sim}$dwe/SRF/).\\
$^{UH}$These optical magnitudes are in the Johnson ($B$) and Cousins ($R$) University of Hawaii filter system.\\
$^\ddag$The NIR magnitudes of these sources were obtained with ISAAC on the VLT.\\
}
\flushright
{\sl \footnotesize continued...}
\end{table}

\addtocounter{table}{-1}
\begin{table}
\footnotesize
\begin{center}
\caption{Follow-up Photometry (ctd)}
\begin{tabular}{lcccccccc}
 Object  & RA$_{\rm opt}$ & Dec$_{\rm opt}$ & B & R & I & J & H & K\\
         &  J2000         &    J2000        &   &   &   &   &   & \\
\hline\hline
A2390\_4 & 21:53:01.0    & 17:44:52 & 22.0$\pm$0.5$^D$ & $>$21.0$^D$  & --  & -- & -- & \\
A2390\_6 & 21:53:20.7    & 17:41:20 & $>$22.5$^D$ &21.96$\pm$0.10 & --  & -- & -- & -- \\
A2390\_8 & 21:53:23.9    & 17:39:30 & 23.69$\pm$0.37 & 21.84$\pm$0.08 & 20.24$\pm$0.06 & 19.50$\pm$0.06 & 18.38$\pm$0.04 & 17.42$\pm$0.03 \\
A2390\_10 & 21:53:25.3    & 17:43:22 & 22.39$\pm$0.09 & 19.91$\pm$0.03 & 19.01$\pm$0.03 & --& -- & --\\
A2390\_12 & 21:53:29.1 & 17:48:55  &   $>$22.5$^D$ & $>$21.0$^D$  &  --  & 19.48$\pm$0.05 & 18.35$\pm$0.03 & 17.70$\pm$0.03\\
A2390\_13 & 21:53:31.6  & 17:47:04  &   $>$22.5$^D$ & $>$21.0$^D$  & -- & -- & -- & -- \\
A2390\_14 & 21:53:32.9    & 17:32:24 & $>$22.5$^D$ &  $>$21.0$^D$ &  --  & 18.32$\pm$0.03 & 17.59$\pm$0.03 & 16.78$\pm$0.02 \\
A2390\_15$^S$ & 21:53:33.2    & 17:42:10 & $>$23.72 & $>$23.25 & $>$22.88$^\circ$ & $>$21.71 & 20.07$\pm$0.06 & 18.83$\pm$0.04 \\
A2390\_16$^{2.7}$ & 21:53:33.6    & 17:37:36 & 22.0$\pm$0.5$^D$ & 20.98$\pm$0.06 &  --  & 19.82$\pm$0.07 & 19.55$\pm$0.10 & 18.90$\pm$0.10 \\
A2390\_17$^{2.7}$ & 21:53:33.7    & 17:41:16 & $>$23.66 & 22.99$\pm$0.24 & 21.11$\pm$0.10$^\circ$ & 19.38$\pm$0.03 & 18.59$\pm$0.03 & 17.58$\pm$0.02 \\
A2390\_18 & 21:53:34.0    & 17:42:42 & $>$23.74 & 23.86$\pm$0.37 & 21.13$\pm$0.11$^\circ$ & 18.65$\pm$0.03 & 17.57$\pm$0.02 & 16.49$\pm$0.02\\
A2390\_19 & 21:53:34.9    & 17:36:32 & 20.50$\pm$0.30$^A$ & 19.35$\pm$0.03 &  --  & 18.74$\pm$0.03 & 17.80$\pm$0.03 & 17.37$\pm$0.03\\
A2390\_20 & 21:53:40.8    & 17:44:16 & 21.90$\pm$0.07 & 20.18$\pm$0.03 & 19.38$\pm$0.02 & 18.22$\pm$0.02 & 17.64$\pm$0.02 & 16.43$\pm$0.02\\
A2390\_24 & 21:53:45.6    & 17:41:48 & 20.26$\pm$0.04 & 18.24$\pm$0.02 & 17.41$\pm$0.03 & 16.31$\pm$0.02 & 15.63$\pm$0.02 & 14.90$\pm$0.02\\
A2390\_26 & 21:53:48.0    & 17:37:56 & 20.10$^A$& 19.85$\pm$0.03 & -- & -- & -- & -- \\
A2390\_27$^{2.7}$ & 21:53:48.0    & 17:42:48 & 21.97$\pm$0.09 & 20.91$\pm$0.05 & 19.22$\pm$0.04 & 18.97$\pm$0.04 & 18.42$\pm$0.04 & 17.62$\pm$0.03 \\
A2390\_28 & 21:53:48.9    & 17:42:00 & 22.20$\pm$0.10 & 21.30$\pm$0.08 & 20.58$\pm$0.05 & 18.98$\pm$0.03 & 18.19$\pm$0.03 & 17.21$\pm$0.02\\
\hline\hline 
\end{tabular}
\end{center}
{\sl
~~~\par
$^A$Magntiude estimated from the APM Palomar Observatory Sky Survey.\\
$^\circ$The three $I$-band magnitudes marked thus were measured on a CFHT image.\\
$^{2.7}$A 2.7 arcsec diameter aperture was used for the photometry of these sources, due to the presence of close neighbouring sources.\\
$^S$This source also has {\sl ISO} 6.75- and 15-${\rm{\mu}}$m detections of 110 and 350${\rm{\mu}}$Jy respectively (L\'emonon et al 1998). These translate to magnitudes of 14.8 and 11.9.
}
\end{table}

\begin{table}
\small
\begin{center}
\caption{Redshifts \label{tab:zphot}}
\begin{tabular}{lcccccccr}
Source & \zspec & \zphot & Filters & Galaxy Type & Age & $A_{V}$ & M$_B$ & $z_{\rm secondary}$ ($\Delta\chi^2$)\\
       &        &(90\%)  &         &             &(Gyr)&         &       &  \\
(1)    &   (2)  &  (3)   &   (4)   &   (5)       & (6) &  (7)    &  (8)  & (9)\\
\hline 
\hline
Per\_2        & 1.307 &    --       &  --     &  --   & --  & --  &  --    & -- \\%
\hline 
RXJ0821\_1    & -- & 1.38 (1.0, 1.5) & 5 & Burst & 0.9 & 0.0 & --22.5 & 0.5 (6.9)\\
RXJ0821\_2    & -- & 1.46 (1.2, 2.1) & 5 & Burst & 0.01 & 0.0 & --23.4 & 3.8 (--5.6)\\
RXJ0821\_3    & -- & 0.62 (0.5, 0.9) & 5 & Burst & 0.2 & 0.0 & --21.2 & 2.7 (3.1)\\
RXJ0821\_4    & -- & 0.82 (0.5, 0.9) & 5 &   E   & 0.5 & 0.5 & --22.7 & 3.8 (4.5)\\
\hline
0902\_1       & -- & 1.10 (0.3, 1.3) & 5 &   E   & 0.6 & 0.8 & --21.5 & 5.2 (1.0)\\
\hline
IRAS09\_1     & -- & 1.10 (0.8, 1.9) & 5 & Burst & 0.5 & 0.9 & --22.6 & 5.2 (0.1)\\
IRAS09\_2     & -- & 1.50 (0.9, 2.3) & 5 &   E   & 4.2 & 0.9 & --23.0 & 3.4 (--0.4)\\
IRAS09\_3     & -- & 1.43 (0.7, 2.1) & 5 & Burst & 0.6 & 0.4 & --24.2 & 0.3 (1.2)\\
\hline
A963\_4       & -- & 0.89 (0.3, 1.0) & 6 & Burst & 0.4 & 1.5 & --21.0 & 5.1 (0.1))\\
A963\_6       & -- & 1.23 (1.0, 1.4) & 6 & Burst & 0.4 & 0.3 & --22.7 & 3.9 (1.7)\\
A963\_7       & -- & 0.40 (0.3, 0.5) & 7 & Burst & 0.7 & 0.2 & --20.3 & 2.8 (--0.1)\\
A963\_8       & -- & 1.03 (0.9, 1.3) & 6 & S0 & 5.6 & 0.8 & --21.3 & 2.4 (4.2)\\
A963\_10      & -- & 0.46 (0.3, 0.8) & 6 & Burst & 0.7 & 0.0 & --23.9 & 3.3 (0.4)\\
A963\_13      & -- & 0.86 (0.7, 1.1) & 6 & Burst & 0.2 & 0.9 & --21.9 & 3.4 (--6.6)\\
A963\_15      & 0.536?$^{L1}$ & 0.56 (0.4, 0.7) & 7 & Burst & 0.5 & 0.9 & --22.1 & 3.4 (8.5)\\
A963\_17      & -- & 0.83 (0.4, 0.9) & 6 & Burst & 0.4 & 0.7 & --22.9 & 3.6 (--7.3)\\
A963\_20      & -- & 2.40 (2.0, 2.8) & 5 & QSO & -- & 0.6 & --24.1 & 0.1 (--0.6)\\
\hline 
A1795\_1      & -- & 1.99 (0.8, 2.5) & 5 & CWW Sbc & -- & 0.6 & --21.5 & 3.3 (--0.6)\\%
A1795\_2      & -- & 1.30 (0.8, 2.0) & 5 & Burst & 0.4 & 0.8 & --23.9 & 4.7 (1.0)\\%
\hline
3C294\_1   & -- & 0.42 (0.2, 0.7) & 6 & Burst & 1.0 & 0.8  & --21.3 & 3.4 (9.9)\\%
3C294\_2   & -- & 3.32 (1.4, 5.4) & 6 & Burst & 0.1 & 0.0  & --22.9 & 0.6 (0.6)\\%
3C294\_3   & -- & 1.94 (0.9, 2.0) & 5 &   E   & 3.3 & 0.0  & --22.2 & 3.0 (1.2)\\%
\hline
A\,1835\_1    & 1.256? & 1.23 (1.0, 1.5) & 5 & Burst & 0.5 & 0.2 & --21.5 & 4.6 (1.8)\\
A\,1835\_2    & 3.830? & 3.20 (2.9, 4.0) & 6 & QSO   & --  & 0.7 & --25.0 & 3.6 (--5.4)\\
A\,1835\_3    & -- & 0.87 (0.7, 1.0) & 6 & E & 2.0 & 0.9 & --22.7 & 4.8 (3.1)\\
A\,1835\_4    & $\approx$0.25$^{P1}$ & 0.71 (0.3, 1.0) & 6 & E & 1.1 & 0.9 & --24.9 & 3.3 (--2.2)\\
\hline
\end{tabular}
\end{center}
{\sl\footnotesize
~~\par
$^{(4)}$\lq Filters\rq\ in column 4 shows the numbers of filters used for the \zphot\ fit.\\
$^{(8)}$M$_B$ in column 8 is the absolute Vega Magnitude in the {\sl
B} Bessell filter. Those sources with a magnitude marked by a $\ddag$
have been corrected for the magnification due to gravitational
lensing.\\
$^{(9)}$The $\Delta\chi^2$ in the last column is the change from the primary to the secondary solution. See \citet{thesis} for more details.\\
$^{L1}$Tentative redshift identification by \citet{lavery93}.\\
$^{P1}$Identified with source 41, XMMUJ140145.0$+$025330 of \citet{piconcelli02}. They quote two possibilities for the redshift, both close to 0.25.\\
}
\flushright
{\sl \footnotesize continued...}
\end{table}

\addtocounter{table}{-1}
\begin{table}
\small
\begin{center}
\caption{Redshifts (ctd)}
\begin{tabular}{lcccccccr}
Source & \zspec & \zphot & Filters & Galaxy Type & Age & $A_{V}$ & M$_B$ & $z_{\rm secondary}$ ($\Delta\chi^2$) \\
       &        &(90\%)  &                    &             &(Gyr)&         &       & \\
(1)    &   (2)  &  (3)   &   (4)   &   (5)       & (6) &  (7)    &  (8)  & (9)\\
\hline 
\hline 
A2199\_1      & -- & 1.01 (0.9, 1.1) & 6 & Burst & 0.6 & 0.2 & --23.3 & 0.8 (10.0)\\
A2199\_2      & -- & 2.83 (1.0, 3.4) & 6 &  QSO  & --  & 0.1 & --22.9 & 0.6 (--1.0)\\
A2199\_3      & -- & 1.73 (1.0, 2.7) & 6 & Burst & 0.4 & 0.7 & --22.6 & 3.1 (4.5)\\
A2199\_4      & -- & 1.20 (0.8, 2.7) & 6 &   E   & 1.2 & 0.9 & --21.0 & 2.6 (0.9)\\
A2199\_5      & -- & 0.54 (0.4, 0.7) & 6 &   S0  & 7.7 & 0.9 & --20.1 & 1.0 (2.9)\\
A2199\_6      &0.029$^{N1}$& 0.06 (0.0, 0.1) & 6 & Burst & 0.7 & 0.0 & --21.7 & 0.5 (--12.8)\\
A2199\_7      & -- & 0.85 (0.8, 0.9) & 6 & Burst & 6.4 & 0.9 & --21.0 & 2.6 (18.5)\\
A2199\_8      & -- & 0.54 (0.4, 0.7) & 6 &   E   & 0.5 & 0.1 & --22.6 & 2.7 (--5.9)\\
A2199\_9      & -- & 0.65 (0.4, 0.9) & 6 &   Sc  & 7.4 & 0.9 & --21.3 & 2.4 (--3.2)\\
\hline
A\,2204\_1    & -- & 0.49 (0.4, 0.6) & 5 & Burst & 0.5 & 0.9 & --22.3 & 3.4 (--10.2)\\
A\,2204\_2    & -- & 0.45 (0.1, 0.9) & 5 & Burst & 0.4 & 0.9 & --21.4 & 3.4 (--0.4)\\
\hline
MS2137\_1     & 2.176 &     --        &   --    &  --   & --  &  -- &   --   &  --\\
MS2137\_3     & -- & 0.70 (0.6, 0.9) & 6 & Burst & 0.8 & 0.6 & --21.6 & 3.7 (4.4)\\
MS2137\_4     & -- & 3.33 (3.1, 3.4) & 5 & Burst & 0.4 & 0.0 & --24.7 & 4.1 (13.9)\\
\hline
A2390\_4      & 1.2172         & -- & -- & -- & -- & -- & -- \\
A2390\_6      & 1.0690         & -- & -- & -- & -- & -- & -- \\
A2390\_8      & 0.8030         & 0.79 (0.7, 0.8) & 6 & Burst & 0.4 & 0.6 & --22.3 & 4.3 (17.0) \\
A2390\_10     & 0.2242         & -- & -- & -- & -- & -- & -- \\
A2390\_12     & --             & 2.28 (1.7, 3.3) & 5 & Burst & 0.3 & 0.3 & --25.0 & 0.6 (0.1) \\
A2390\_13     & 1.5933         & -- & -- & -- & -- & -- & -- \\
A2390\_14     & --             & 1.35 (0.9, 1.8) & 5 & Burst & 0.9 & 0.0 & --24.5 & 5.3 (--0.1)\\
A2390\_15     & -- $^{C3}$     & 2.78 (2.1, 3.3) & 7 & E     & 1.8 & 1.8  &  --21.9$^\ddag$ & 1.1 (4.7) \\
A2390\_16     & 1.6321         & 1.35 (1.2, 2.1) & 5 & QSO   & --  & 0.3& --23.3 & 0.4 (1.5) \\
A2390\_17     & 1.466 $^{C2}$  & 1.11 (1.0, 1.2) & 6 & Burst & 0.6 & 0.6    & --21.8$^\ddag$ & 4.6 (8.6) \\
A2390\_18     & 1.467 $^{C1}$  & 1.45 (1.3, 1.5) & 6 & Burst & 3.3 & 0.0 & --23.2$^\ddag$ & 0.4 (29.9) \\
A2390\_19     & 1.6750         & 0.31 (0.2, 0.4) & 5 & QSO   & --   & 0.6 & --21.6 & 0.9 (2.0) \\
A2390\_20     & 0.305          & 0.22 (0.2, 0.4) & 6 & CWW Scd with AGN & -- & 0.6 & --19.4 & 3.1 (16.5)  \\
A2390\_24$^*$ & 0.214          & 0.26 (0.2, 0.5) & 6 & E     & 4.6  & 0.0 & --21.8 & 0.7 (12.0)\\
A2390\_26     & 1.5187         & -- & -- & -- & -- & -- & -- \\
A2390\_27     & --             & 0.87 (0.8, 0.9) & 6 & Burst & 0.2 & 0.0 & --23.6 & 2.0 (35.8)\\
A2390\_28     & 1.0330         & 1.21 (1.2, 1.3) & 6 & S0    & 0.9 & 0.9 & --23.6 & 0.4 (13.3)\\
\hline 
\end{tabular}
\end{center}
{\sl\footnotesize
~~\par 
$^{N1}$Identified with MCG$+07-34-048$. Source: NASA Extragalactic Database online; \citet{mcg07-34-048}.\\
$^{C1}$ This is a spectroscopic redshift from Cowie et al
(2001). \\ 
$^{C2}$ This is a spectroscopic redshift from Cowie et al
(2001) based on the single identification of a line, but also
consistent with their photometric redshift estimate of 1.5$\pm0.2$. A
different identification would imply a redshift of 2.317 instead. \\
$^{C3}$ Cowie et al (2001) obtain a photometric redshift of 2.6$_{-0.2}^{+0.1}$.\\ 
$^*$ This source is
identified as an Sbc galaxy with a redshift of $z=0.21501$ by Yee et
al (1991). A fit to an Sbc template gives $z_{\rm phot}=0.3$.\\ 
}
\end{table}



\end{landscape}


\end{document}